\documentclass[reprint,amsmath,amssymb,aps,prb]{revtex4-2}
\usepackage{graphicx}           
\usepackage{dcolumn}            
\usepackage{bm}                 
\usepackage[hypertex]{hyperref} 
\usepackage{xcolor,fancybox}    %
\newcommand{\bra}[1]{\ensuremath{\bm{\langle}#1\bm{|}}}
\newcommand{\ket}[1]{\ensuremath{\bm{|}#1\bm{\rangle}}}

\newcommand{\Av}[1]{\ensuremath{\big<#1\big>}}
\newcommand{\AV}[1]{\ensuremath{\Big<#1\Big>}}

\begin{document}

\title{The effect of Anderson localization on surface plasmon polariton propagation and outward leakage when scattered by a randomly corrugated section of the interface}


\author{Yu.\,V.~Tarasov}
\email[Corresponding author: ]{yuriy.tarasov@gmail.com}

\author{O.\,M. Stadnyk}
\affiliation{O.~Ya.~Usikov Institute for Radiophysics and Electronics of the National Academy of Sciences of Ukraine,\\
12 Acad.~Proskura~St., Kharkiv 61085, Ukraine}

\date{\today}

\begin{abstract}
The practical applications of surface plasmon polaritons (SPPs) require the deep understanding of the impact of electrical characteristics variability and geometrical irregularity of the metal-dielectric interface. Traditional methods in the theory of wave scattering at rough interfaces fail to treat simultaneously the interference (Anderson) localization of the SPP, which may arise due to its multiple back-scattering by random distortions of surface relief, and the leakage into the uniform dielectric half-space. In our previous works [Low Temp. Phys. \textbf{42}, 685 (2016); Ann.~Phys.~\textbf{455}, 169378 (2023)], by representing the perturbation of surface impedance as an effective potential in the Schr\"odinger-like equation, we suggested the way to describe the interplay between Anderson localization and the leakage of the SPP. In the present study we show that the problem of SPP scattering from finite geometrically rough region of the interface can be reduced to the problem of its scattering from the same region but with effective random impedance. We calculate the radiation pattern and demonstrate its pronounced anisotropy that arises due to the interplay between different geometrical parameters of the interface roughness.
\end{abstract}
%
\maketitle

\section{Introduction}
%

In recent years, there has been an explosion of interest in controlling the light-metal interaction in nanostructures with geometric length smaller than the light wavelength \cite{Nature424_824, MaierAtwater05, Ebbesen98, Lezec02,PhysRevLett.90.213901,Lockyear_2005}. These studies aim to develop various applications as well as to understand the fundamentals of electromagnetic interaction between optical nano-objects, where scattering of surface plasmon polaritons (SPPs) and their transformation into bulk waves are the determining factors. Methods for describing the SPP transformation into bulk waves upon scattering by obstacles differ substantially for surfaces with impedance and relief inhomogeneities, and there is still no common view on this problem.

Although the scattering of SPP by regular and random inhomogeneities of the surface impedance of metals was studied by many authors (see, for example, \cite{FreiYurk93, PhysRevB.62.6228, Nikitin07, PhysRevB.77.195441, TarUsIak16, TarStadKvit23}), studies of scattering on inhomogeneous random \textit{relief} of the surface~\cite{PhysRevLett.54.1559, Maystre:94, Maradudin:95, PhysRevB.57.4132, MarFreiSimLesk01, Sanchez-Gil:02, Shchegrov2004MultipleSO, Zayats05} has even longer history.
The goals of these studies were to reveal the potential Anderson localization of SPP due to its multiple back-scattering by random inhomogeneities of the interface and to analyze the outflow of the SPP energy into the free space.

The above mentioned goals, at first glance, may seem to be poorly compatible since the localization of Anderson nature is a subtle interference effect extremely sensitive to violation of the coherence of the wave during its multiple scattering. Meanwhile, the metal surface with finite-size scatterers located on it is an essentially non-Hermitian system, for which it is rather problematic to assume the scattering coherence. Scattering in non-Hermitian systems is a relatively new area of research~\cite{Moiseyev2011}, and sophisticated effects in such systems, in particular the interference-induced Anderson localization, still await a convincing explanation.

In our paper \cite{TarStadKvit23}, we have suggested an efficient method for solving the problem of simultaneous localization and leakage of the SPP running into a finite area of the surface with randomly inhomogeneous impedance. The main idea of the proposed method is to represent the impedance perturbation in terms of some \textit{effective potentials} in the governing Schr\"odinger-like equation. We have shown that upon scattering of the SPP by nonuniform boundary segment the combined waves are excited, which we referred to as \textit{composite} plasmon polaritons \cite{TarUsIak16,TarStadKvit23}.

The particular property of the SPP scattering in waveguide-type systems with corrugated boundaries is the formation in them of so-called \textit{entropy barriers}~\cite{Zwanzig92}. The scattering from such barriers is determined by not the relative amplitude of the roughness, but rather by its gradient, i.\,e., by the degree of its smoothness~\cite{TarShost15}. In the present paper, we consider a wave system that is in a sense also of waveguide nature, yet \textit{semi}-confined in the transverse direction. There is no competition in it between the discovered in \cite{bib:MakTar98,bib:MakTar01} \textit{by-height} and \textit{by-slope} scattering mechanisms, named later on more appropriately in~\cite{Izrailev2003} the \textit{amplitude} and the \textit{gradient} mechanisms. The primary goal of this work is to develop a method for investigating the properties of the SPP scattering by random roughness of metal surface, taking account of the entropy barrier formed in its finite segment and the gradient mechanism of surface scattering, which is shown to be in our problem non-alternative.

%
\section{Problem Statement}
\label{Formulation}
\begin{figure}[h]
  \includegraphics[width=8.6cm]{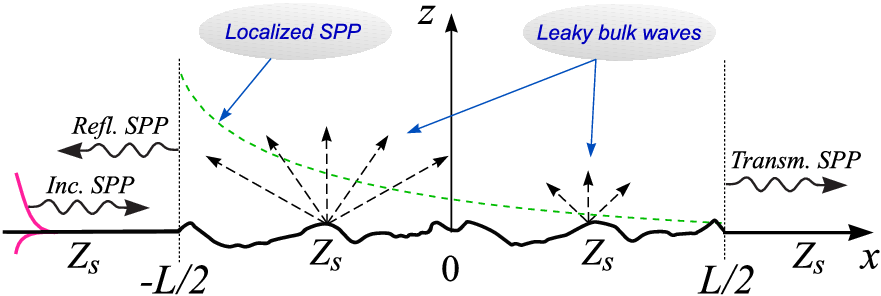}
  \caption{Geometry of SPP scattering by a~ran\-domly corrugated boundary segment of finite length~$L$.
  \hfill\label{fig1}}
\end{figure}

%
Consider the two-dimensional problem of scattering of a surface plasmon polariton excited by some source (e.\,g., a~slit \cite{bib:Tejeira05,bib:Tejeira07}) on the boundary $\bm{S}$ specified by formula
\begin{equation}\label{Surface_equation}
  \bm{S}\!: z=
\begin{cases}
 \xi(x) & \text{if}\quad x\in\mathbb{L}\ ,\\
 \ \ {0} & \text{if}\quad x\not\in\mathbb{L}\ .
\end{cases}
\end{equation}
Symbol $\mathbb{L}$ denotes a rough segment of the boundary $\big(|x|<L/2\big)$, $\xi(x)$ is the arbitrary single-valued random function of coordinate $x$ (Fig.~\ref{fig1}). Surface impedance at all points of the uneven boundary is assumed to have the same (complex) value.
The only nonzero SPP magnetic field component $H(\mathbf{r})\equiv H_y(\mathbf{r})$, where $\mathbf{r}=(x,z)$ is the two-dimensional radius vector, satisfies Helmholtz equation
\begin{subequations}\label{initial_eq_Psi+BC}
\begin{equation}\label{initial_eq_Psi}
   \left(\Delta+k^2\right)H(\mathbf{r})=0
\end{equation}
and impedance boundary condition (BC) \cite{Jackson99}
\begin{equation}\label{Surf_imp-def}
  \mathbf{E}_{\parallel}(x,z\in \bm{S})=Z_s\big[\mathbf{n}\times\mathbf{H}(x,z\in \bm{S})\big]\ ,
\end{equation}
\end{subequations}
Here $\mathbf{E}_{\parallel}(x,z\in \bm{S})$ is the tangential electric field component, the complex surface
impedance ${Z_s=Z'_s+iZ''_s }$ is assumed to be constant-valued at all points of the uneven boundary, $\mathbf{n}$ is the unit vector of the \textit{outer} normal to it; $\Delta$~is the (two-dimensional) Laplace operator, $k=\omega/c$ is the wavenumber in vacuum.

In what follows we will use the BC at the metal surface in a different yet equivalent form, specifically,
\begin{align}\label{Initial_BC}
  \left.\left(\frac{\partial H}{\partial n}+ikZ_s H\right)\right|_{\mathbf{S}}=0\ .
\end{align}
Here, $\partial/\partial n$ is the derivative with respect to the outer normal to surface $\mathbf{S}$. The condition for applicability of BC \eqref{Initial_BC} (see Ref.~\cite{GarciaMaradLesk:90}) are assumed to be fulfilled.

To solve equation \eqref{initial_eq_Psi} it is necessary to set also the boundary conditions at both ends of the rough interval~$\mathbb{L}$, which depend on physical formulation of the problem, and at an infinite distance along $z$ coordinate. In fact, we will solve the \emph{scattering} problem for SPP incident onto an inhomogeneous segment from the left, see Fig.~\ref{fig1}, and scattered by this segment in all directions. At large distances from the scattering region ($\mathbb{L}$) the system openness requires the fulfillment of Sommerfeld's radiation condition~\cite{Schot1992}: the field scattered into free volume must contain outgoing harmonics only. As for the \emph{total} field, which includes also the field of the incident wave, the conditions for it must be formulated taking account of the nature and the location of primary radiation sources.

In the case of an ideally flat boundary and under condition $Z''_s<0$ (the absorbing metal), the solution to problem \eqref{initial_eq_Psi+BC} consists of two surface plasmon-polaritons of the following form,
\begin{align}\label{RightLeft_SPP}
 & H_{\text{spp}}^{(\pm)}(\mathbf{r})= \mathcal{A}^{(\pm)}
  \exp\big(\pm ik_{\text{spp}}x-iZ_s kz\big)\ ,\\
\notag
 & k_{\text{spp}} = k'_{\text{spp}} + ik''_{\text{spp}} = k\sqrt{1-Z_s^2}\ .
\end{align}
We will be interested in the SPPs propagating along a smooth boundary over distances much larger than their wavelength. For this, from \eqref{RightLeft_SPP} it follows that the inequality must be met
\begin{subequations}\label{SPP}
\begin{equation}\label{freeSPP}
  |k''_{\text{spp}}| \ll k'_{\text{spp}}\ .
\end{equation}
In metals, this usually takes place at not too large frequencies (see, for example, Ref.~\cite{Palik98}). From inequality \eqref{freeSPP} it follows that
\begin{equation}\label{Low_dissip}
  Z'_s\ll |Z''_s|\ ,
\end{equation}
\end{subequations}
and the SPP propagation constant is, consequently,
\begin{equation}\label{SPP_asymp_wavenumber}
  k_{\text{spp}} \approx k\left(\sqrt{1+{(Z''_s)}^2}
  + i\frac{Z'_s|Z''_s|}{\sqrt{1+{Z''_s}^2}}\right)\ .
\end{equation}
Based on \eqref{SPP_asymp_wavenumber}, we may introduce the SPP dissipative attenuation length along the metal-vacuum interface, viz.,
\begin{equation}\label{L(SPP)_dis}
  L^{\scriptscriptstyle{(\text{SPP})}}_{dis}= \big|k''_{\text{spp}}\big|^{-1}
  = \frac{\sqrt{1+{Z''_s}^2}}{kZ'_s|Z''_s|}\ .
\end{equation}
%
%
\section{The method for solving the SPP scattering problem}
\label{Trial_solution}
%
The main idea of the approach we propose is to look for a solution to problem \eqref{initial_eq_Psi+BC} as a sum
\begin{align}\label{h-BC}
  H(\mathbf{r})=H_0(\mathbf{r})+h(\mathbf{r})\ ,
\end{align}
where function $H_0(\mathbf{r})$ is taken as a ``trial'' field of purely plasmon-polariton type and $h(\mathbf{r})$ to be the solution portion containing ``leaky'' harmonics with non-zero normal-to-the-surface wave vector components.

Trial function $H_0(\mathbf{r})$, which by assumption does not contain up-leaking harmonics, will be modeled in the form conventional for \textit{one-dimensional} scattering problems. Specifically, in the left ($l$) region ${x<-L/2}$ in Fig.~\ref{fig1} it will be sought as a sum of incident SPP which we choose, for convenience, to have unit amplitude at the point of incidence onto segment $\mathbb{L}$, and a reflected SPP:
\begin{subequations}\label{H_0^lrc}
\begin{align}\label{H_0^<}
  H_0^{(l)}(\mathbf{r})=& \Big\{ e^{ik_{\text{spp}}\big(x+ L/2\big)}
  + r_- e^{-ik_{\text{spp}}\big(x+ L/2\big)} \Big\} e^{-iZ_s kz}\ .
\end{align}
In the right ($r$) region $x>L/2$, we choose field $H_0(\mathbf{r})$ in the form of \emph{transmitted} SPP wave, namely,
\begin{align}\label{H_0^>}
  H_0^{(r)}(\mathbf{r})=t_+e^{ik_{\text{spp}}\big(x- L/2\big)-iZ_s kz}\ .
\end{align}
\end{subequations}
In \eqref{H_0^lrc}, factors $r_-$ and $t_+$ play the role of reflection and transmission coefficients for SPP incident onto defective segment of the boundary.

The choice of the trial field inside the intermediate region $x\in\mathbb{L}$ ($in$-region) requires especial consideration. To obtain the proper model for this field, we solve the auxiliary problem of the SPP propagation along the infinitely long corrugated surface. For this, we, first, flatten the rough metal boundary using the following coordinate transformations,
\begin{subequations}\label{Smoothing_transforms}
\begin{align}
\label{Smoothing_transform-x}
  x & =x'\ , \\
\label{Smoothing_transform-z}
  z & =z'+\xi(x')\theta(L/2-|x'|)\ .
\end{align}
\end{subequations}
Under conditions $\xi(\pm L/2) = 0$ and $\xi'(\pm L/2) = 0$ equalities \eqref{Smoothing_transforms} imply the following derivative change,
\begin{subequations}\label{Derivative_transforms}
  \begin{align}
\label{d/dx}
   \frac{\partial}{\partial x} & =\frac{\partial}{\partial x'}-\theta(L/2-|x'|)\xi'(x')\frac{\partial}{\partial z'}\ ,\\
\label{d/dz}
   \frac{\partial}{\partial z} & =\frac{\partial}{\partial z'}\ .
  \end{align}
\end{subequations}
Hereby we reduce the Laplace operator in \eqref{initial_eq_Psi+BC} to the sum
\begin{align}\label{Laplacian'}
  \Delta=& \left(\frac{\partial^2}{\partial {x'}^2}+\frac{\partial^2}{\partial {z'}^2}\right)+
  {\xi'}^2(x')\frac{\partial^2}{\partial {z'}^2}
\notag\\
 &\ -\left[\xi'(x')\frac{\partial}{\partial x'}
  +\frac{\partial}{\partial x'}\xi'(x')\right]\frac{\partial}{\partial z'}\ .
\end{align}
where the first term on the right-hand side is the unperturbed Laplacian and the ``perturbation'' has the form of two fundamentally different operator terms, namely, quadratic and linear in derivative $\xi' (x)$. From \eqref{Laplacian'} it can be seen that the scattering caused by the rough part of the boundary is determined by the roughness ``sharpness'' (first derivative $\xi' (x)$) and its local curvature (second derivative $\xi''(x)$), but not directly by ``height'' function~$\xi(x)$.

The transform \eqref{Smoothing_transforms} affects, in addition to equation~\eqref{initial_eq_Psi}, also boundary condition \eqref{Initial_BC}. Considering that in the original coordinate system $\mathbf{K}$ the derivative along the \textit{local} normal to the metal surface is equal to
\begin{align}\label{Normal_deriv_K}
  \frac{\partial H(\mathbf{r})}{\partial n}=
  \frac{1}{\sqrt{1+\left[\xi'(x)\right]^2}}
  \left(-\xi'(x)\frac{\partial H}{\partial x} +\frac{\partial H}{\partial z} \right)\ ,
\end{align}
in a coordinate system with new, primed variables (we refer to it as $\mathbf{K'}$), the BC \eqref{Initial_BC} acquires the form
\begin{align}\label{Local_BC_K'}
  \Bigg\{ & \frac{-\xi'(x')}{\sqrt{1+\left[\xi'(x')\right]^2}}\frac{\partial}{\partial x'}
  +\sqrt{1+\left[\xi'(x')\right]^2}\frac{\partial}{\partial z'} + ikZ_s\Bigg\} \times
\notag\\
 &
 \times H\big[\mathbf{r}(x'z')\big]\bigg|_{z'=0}=0\ .
\end{align}
Condition \eqref{Local_BC_K'} is difficult to work with, so we will use it in limiting cases of smooth ($\Av{\left[\xi'(x')\right]^2}\ll 1$) and sharp ($\Av{\left[\xi'(x')\right]^2}\gg 1$) boundary roughness. In the first case, the BC reduces to the approximate form
\begin{subequations}\label{NEWlocal_BCs}
\begin{align}\label{Local_BC_smooth}
  \left.\left(\frac{\partial H}{\partial z'}
  +ikZ_s H\right)\right|_{z'=0}\approx 0\ ,
\end{align}
and in the second case to equality
\begin{equation}\label{Local_BC_sharp}
  \left.\left( \frac{\partial H}{\partial z'}+ ik\frac{Z_s}{\sqrt{1+\left[\xi'(x')\right]^2}} H\right)\right|_{z'=0}\approx 0\ .
\end{equation}
\end{subequations}
The functional structure of equalities \eqref{NEWlocal_BCs} is such that hereinafter we can use the equality \eqref{Local_BC_sharp} as a universal BC for both smooth and sharp roughnesses. In the latter case we only must have in mind that the limitations imposed by the applicability of local BC \eqref{Local_BC_sharp} restrict the admissible upper value of the corrugation sharpness.

Below, to make the expressions more clear  we will not specify prime index for the variables of $\mathbf{K}'$ system, if it does not cause misapprehension. Taking into account the presence of terms quadratic in $\xi'(x)$ both in \eqref{Laplacian'} and in \eqref{Local_BC_sharp}, we introduce, in addition to the original random function $\xi' (x)$, one more random functional variable $\Delta[\xi'(x)]^2=[\xi'(x)]^2-\sigma^2$, where $\sigma^2=\Av{ [\xi'(x)]^2}$, angle brackets denote statistical averaging. Both of these random functions have zero mean values, which makes them convenient functional variables for further averaging procedure. Using these two functional variables, the Helmholtz equation with Laplacian \eqref{Laplacian'} and boundary condition \eqref{Local_BC_sharp} can be rewritten as the following boundary value problem:
\begin{subequations}\label{Statement-fin}
\begin{align}
\label{Fin_eq}
 & \left\{\left[\frac{\partial^2}{\partial x^2}+(1+\sigma^2)\frac{\partial^2}{\partial z^2}+k^2\right]-
  \hat{V}(x)\right\} H = 0 \ ,\\[6pt]
\label{Fin_BC}
 & \Biggl\{\frac{\partial H}{\partial z} + ik \frac{Z_s}{\sqrt{1+\sigma^2+\Delta\left[\xi'(x)\right]^2}}H\Biggr\}\Biggr|_{z=0}=0\ .
\end{align}
\end{subequations}
In \eqref{Fin_eq} we have introduced, by analogy with Schr\"odinger equation, the effective random ``potential''
\begin{subequations}\label{Dyn_Pots}
\begin{align}
\label{Dyn_Pot_V}
 & \hat{V}(x)=\hat{V}_1(x)+\hat{V}_2(x)\ ,\\
\label{Dyn_Pot_V1}
 & \hat{V}_1(x)=-\Delta\big[{\xi'}(x)\big]^2\frac{\partial^2}{\partial z^2}\ ,\\
\label{Dyn_Pot_V2}
 & \hat{V}_2(x)=\left[\xi'(x)\frac{\partial}{\partial x}+\frac{\partial}{\partial x}\xi'(x)\right]\frac{\partial}{\partial z}\ ,
\end{align}
\end{subequations}
which has an operator structure. In BC \eqref{Fin_BC}, it is reasonable to introduce the \textit{effective} surface impedance
\begin{align}\label{Z^(eff)}
  Z_s^{\text{(eff)}}(x)=\frac{Z_s}{\sqrt{1+\sigma^2+\Delta\left[\xi'(x)\right]^2}}\ ,
\end{align}
which, in contrast to the initial uniform impedance $Z_s$, depends on coordinate $x$.

The problem \eqref{Statement-fin} can be regarded as consisting of two sub-problems of fundamentally different types. The first sub-problem is to take properly into account the \textit{one-dimensional} (1D) nature of random potentials $\hat{V}_1(x)$ and $\hat{V}_2(x)$ entering into equation \eqref{Fin_eq} through \eqref{Dyn_Pot_V}. The second sub-problem is to account for \textit{randomness} of the effective impedance in BC \eqref{Fin_BC}. The scattering by random potentials depending on one coordinate only is known for the fundamental fact that, for arbitrarily small values of these potentials, it causes the Anderson-type localization of all (``single-particle'') eigenstates of the system (see, e.\,g., Refs.~\cite{Berez74,AbrikRyzh78,GorDorPrig83,KanerCheb87,FreiTar91,FreiTar01}). The non-uniformity of the surface impedance results in the leakage of SPP waves, which in the absence of inhomogeneity would propagate along one coordinate axis \cite{TarStadKvit23}, into the surrounding space. Solving the problem \eqref{Statement-fin} allows consideration of these two fundamentally different effects, localization and leakage, conjointly.

The approach we suggest is as follows. First, we give an algorithm for finding the leaking field without detailing the spatial structure of the \textit{surface} component of the full-wave field inside the rough boundary segment. Then, following the technique developed in Ref.~\cite{TarStadKvit23}, we find the \textit{radiation} field $h(\mathbf{r})$ starting with the expression for the trial field suitably \textit{modeled} on the entire $x$-axis. At the first stage, we set potentials $\hat{V}_1(x)$ and $\hat{V}_2(x)$ equal to zero, assuming that the scattering produced by them is weak (the weakness criteria will be discussed below). In the regions external to segment $\mathbb{L}$, i.e., for $|x|>L/2$, we have previously chosen the trial field in the form \eqref{H_0^<} and \eqref{H_0^>}. As for the \textit{in}-part of this field, $H_0^{(in)}(\mathbf{r})$, we will seek it in the form
\begin{subequations}\label{H_0(in)}
\begin{align}\label{in_trial field}
   H_0^{(in)}(\mathbf{r})= \mathcal{B}(x)\exp\left(-ik\overline{Z}_s^{(0)}z\right)\ ,
\end{align}
where
\begin{align}\label{Zs(0)}
  \overline{Z}_s^{(0)}=\frac{Z_s }{\sqrt{1+\sigma^2}}
\end{align}
\end{subequations}
is effectively renormalized unperturbed surface impedance.
Formula \eqref{in_trial field} is valid at distances from the ends of segment $\mathbb{L}$ exceeding the correlation radius $r_c$ which will be considered small as compared to the length of the rough boundary segment \cite{MaraduMichel90}. At distances smaller than $r_c$ from the endpoints of the segment, one can no longer neglect the fact that impedance \eqref{Z^(eff)} goes at these points to its basic value $Z_s$ continuously.

%
\section{Calculation of radiation component of the scattered field}
\label{Rad_field}
%
Following \cite{TarUsIak16,TarStadKvit23}, we will seek function $h(\mathbf{r})$ which contains not only surface harmonics, as an integral
\begin{equation}\label{Full_solution_h(r)}
  h(\mathbf{r})=\int\limits_{-\infty}^{\infty}\frac{dq}{2\pi}
  \widetilde{\mathcal{R}}(q)\exp\left[iq x+i\big(k^2-q^2\big)^{1/2}z\right]\ ,
\end{equation}
containing bulk harmonics ``leaking'' from the surface into the free (upper) half-space. We consider the trial field given by equalities \eqref{H_0^lrc} and \eqref{in_trial field} to satisfy Helmholtz equation with Laplacian \eqref{Laplacian'} in that half-space. As for the field \eqref{in_trial field}, we will restrict ourselves to searching for it only in the thin near-surface layer $z\ll k^{-1}$ in order to avoid purely technical problems that may arise when matching the fields at both sides of vertical axes passing through end points $x_{\pm}=\pm L/2$ at $z\neq 0$.

In the upper half-space the Helmholtz equation is simultaneously satisfied by both the trial field and the radiation field $h(\mathbf{r})$. The boundary condition at $z=0$ is specified differently in the three regions of the interface. In the two ``outer'' regions corresponding to $|x|>L/2$ (to be short, we call them \textit{out}-regions), the BC is given by
\begin{equation}\label{BC_lr}
  \left(\frac{\partial H}{\partial z}+ikZ_s H\right)
  \Bigg|_{z=0}=0\ .
\end{equation}
In the region $x\in\mathbb{L}$ (\textit{in}-region), the BC is given by equality \eqref{Fin_BC}, in which, instead of ``bare'' surface impedance $Z_s$, the effective one \eqref{Z^(eff)} appears, which contains the explicit dependence on $x$-coordinate.

Problem \eqref{initial_eq_Psi+BC} for the total field \eqref{h-BC} in the upper half-space with perturbed Laplacian \eqref{Laplacian'} is different in formulation for the \textit{out}- and \textit{in}- regions of the $x$ axis. In the \textit{out}-regions, the Helmholtz equation has the form
\begin{equation}\label{Helmholtz|x|>L/2}
  \left(\frac{\partial^2}{\partial x^2}+\frac{\partial^2}{\partial z^2}+k^2\right)
  \left[H_0(\mathbf{r})+h(\mathbf{r})\right]=0\ .
\end{equation}
The boundary condition in these perfectly flat surface regions is given by \eqref{BC_lr}. In the ``rough'' inner region flattened with transformation \eqref{Smoothing_transforms} the BC is given by equality \eqref{Fin_BC}, which includes an effective non-uniform surface impedance.

After coordinate transformation \eqref{Smoothing_transforms}, the Helmholtz equation on segment $\mathbb{L}$ looks much more complicated than equation~\eqref{Helmholtz|x|>L/2}.
It is made stochastic and not univariate, hence it is not solvable exactly without using some approximations. Instead of solving it directly, we will apply a particular trick which reduces to definition of function $H_0(\mathbf{r})$ \textit{artificially}, being guided, first of all, by considerations of the solution procedure convenience. Thereafter, we find the term $h(\mathbf{r})$ by expressing it, with representation~\eqref{Full_solution_h(r)}, through the beforehand chosen ``fitting'' function~$H_0(\mathbf{r})$.
%
\subsection{Finding the radiative scattering amplitude}
%
Radiation field $h(\mathbf{r})$ over the flattened metal surface satisfies equation \eqref{Helmholtz|x|>L/2} with $H_0^{(l,r)}(\mathbf{r})$ in the \textit{out}-regions of the $x$ axis and equation \eqref{Fin_eq} with $H_0^{(in)}(\mathbf{r})$ from \eqref{in_trial field} in the \textit{in}-region. Since outside the interval~$\mathbb{L}$ the potential $\hat{V}(x)\equiv 0$, we can present Helmholtz equation on the entire $x$ axis in the form
\begin{align}\label{Helmholtz_on_all_x}
  \left\{\frac{\partial^2}{\partial x^2}+\Big[1+\theta\big(L/2-|x|\big)\sigma^2\Big]\frac{\partial^2}{\partial z^2}+k^2-
  \hat{V}(x)\right\}H=0\ .
\end{align}
The boundary condition to this equation on the surface $z=0$ is a combination of three different BCs set in different boundary regions shown in Fig.~\ref{fig1}. For $|x|>L/2$, the BC for the total field has the form \eqref{BC_lr}. The same BC is satisfied by trial fields \eqref{H_0^lrc} chosen before, and hence the same does the radiation field $h(\mathbf{r})$ within \textit{out}-regions of the metal-vacuum interface.

On the ``flattened'' segment~$\mathbb{L}$, the original Helmholtz operator is transformed into the operator in the left-hand side of \eqref{Fin_eq}, and the BC is given by equality \eqref{Fin_BC} containing \textit{roughness-perturbed} effective surface impedance. Since for $|x|>L/2$ we choose the trial solution to have the form of a combination of standard plasmon polaritons~\eqref{H_0^lrc}, function $h(\mathbf{r})$ in these boundary sections is assumed to be identical zero, because with BC \eqref{BC_lr} it must also possess the structure of an unperturbed SPP already included in formulas \eqref{H_0^lrc}. For $|x|<L/2$ we obtain the BC for $h(\mathbf{r})$ by substituting~\eqref{h-BC} into \eqref{Fin_BC} and applying for the trial field in this region the BC following from the beforehand chosen model \eqref{in_trial field}, namely,
\begin{equation}\label{H_0(in)-BC}
  H_0^{(in)}(x,z=0)\equiv\mathcal{B}(x)\ .
\end{equation}
From BC \eqref{Fin_BC} for the entire field in region $\mathbb{L}$ we obtain, taking account of \eqref{in_trial field} and \eqref{H_0(in)-BC}, the following BC equality,
\begin{align}\label{2nd_BC_h(r)}
 \left[\frac{\partial }{\partial z}+ik\overline{Z}_s^{\text{(eff)}}(x)\right] & h(\mathbf{r}) \Bigg|_{z=0}=
\notag\\[3pt]
 & = -ik\left[\overline{Z}_s^{\text{(eff)}}(x)-\overline{Z}_s^{(0)}\right]\mathcal{B}(x)\ .
\end{align}
In the random-impedance problem solved earlier in \cite{TarStadKvit23}, the similar equality has played the role of an equation from which the kernel $\widetilde{\mathcal{R}}(q)$ of the integral \eqref{Full_solution_h(r)} was determined. Introducing notation $\Delta\overline{Z}_s(x)=\overline{Z}_s^{\text{(eff)}}(x)-\overline{Z}_s^{(0)}$, we can rewrite \eqref{2nd_BC_h(r)} in the form
\begin{align}\label{BC_modified}
  \left[\frac{\partial }{\partial z}+ik\overline{Z}_s^{(0)}+ik\Delta\overline{Z}_s(x)\right]h(\mathbf{r})\Bigg|_{z=0}=
  -ik\mathcal{B}(x)\Delta\overline{Z}_s(x)\ ,
\end{align}
which is identical in structure to equation (13) from Ref.~\cite{TarUsIak16}. Substituting $h(\mathbf{r})$ in the form \eqref{Full_solution_h(r)} into \eqref{BC_modified}, then performing Fourier transformation with respect to coordinate $x$ and dividing both sides of the emerging equality by function
\begin{equation}
  Z_{e}\left( q\right) = \overline{Z}_s^{(0)} + \sqrt{1-(q/k)^2}\ ,
\end{equation}
we arrive at the standard Fredholm integral equation of the second kind \cite{bib:KolmogFomin68, bib:Courant_Hilbert66},
\begin{subequations}\label{Fredholm}
\begin{align}\label{Fredholm_eq}
 \notag\displaybreak[0]\\
  \widetilde{\mathcal{R}}(q)+
  \int\limits_{-\infty}^{\infty}\frac{dq'}{2\pi}\mathcal{L}(q,q')\widetilde{\mathcal{R}}(q')
  =-\int\limits_{-\infty}^{\infty}\frac{dq'}{2\pi}
  \mathcal{L}(q,q')\widetilde{\mathcal{B}}(q')\ .
\end{align}
Here, the kernel of the integral operator in the left-hand side has the form
\begin{equation}\label{Kernel_Fredholm_eq}
  \mathcal{L}(q,q') = Z_{e}^{-1}\left( q\right) \Delta\widetilde{\overline{Z}}_s(q-q')\ .
\end{equation}
\end{subequations}
The inverse factor in the right hand side of \eqref{Kernel_Fredholm_eq} does not contain a singularity because the real part of impedance $Z_s$ is nonzero due to the dissipation in metal. The interpretation of this factor in terms of the intermediate scattering modes we have termed \textit{composite plasmons} (CP) was proposed in paper~\cite{TarStadKvit23}.

The formal solution to equation \eqref{Fredholm_eq} can be written in the operator form through the introduction of some operator $\hat{\mathcal{L}}$ matrix elements of which in momentum representation are given by \eqref{Kernel_Fredholm_eq}, viz.,
\begin{equation}\label{FredholmEq-oper}
  \bm{\mathcal{R}}=-\big(\openone+\hat{\mathcal{L}}\big)^{-1}\hat{\mathcal{L}}\,
  \bm{\mathcal{B}}\ .
\end{equation}
Symbols $\bm{\mathcal{R}}$ and $\bm{\mathcal{B}}$ here denote Hilbert space vectors written in any representation appropriate for the solution. Specifically, in the momentum representation equality \eqref{FredholmEq-oper} takes the form
\begin{equation}\label{FredholmSol-momentum}
  \widetilde{\mathcal{R}}(q)=
  -\int\limits_{-\infty}^{\infty}\frac{dq'}{2\pi}\bra{q}\big(\hat{\openone}+\hat{\mathcal{L}}\big)^{-1}
  \hat{\mathcal{L}}\ket{q'}\widetilde{\mathcal{B}}(q')\ ,
\end{equation}
where we use Dirac notations $\bra{\cdot}$ and $\ket{\cdot}$ for \textit{bra}- and \textit{ket}-vectors.
%
\subsection{Determining factor $\bm{\mathcal{B}(x)}$ in trial field~(\ref{in_trial field})}
%
On the interval $\mathbb{L}$, function $H_0(\mathbf{r})$ will be considered to satisfy differential equation \eqref{Helmholtz_on_all_x}. At the same time, we \textit{declaratively} choose the part of this function that depends on coordinate $z$ in the form of a factor suggested in~\eqref{in_trial field}. Such an artificial ``$z$-determinization'' gives an opportunity to get rid of derivatives with respect to $z$ in the definition of potentials \eqref{Dyn_Pots}, replacing them with other simulated potentials which depend on coordinate $x$ only. As a result, for function $\mathcal{B}(x)$ the \textit{univariate} differential equation results
\begin{equation}\label{B(x)-eq}
  \left[\frac{\partial^2}{\partial x^2}+k^2_{\text{spp}}-
  \hat{\mathcal{V}}(x) \right]\mathcal{B}(x)=0\ ,
\end{equation}
where ``potential'' $\hat{\mathcal{V}}(x)$ is represented by expressions
\begin{subequations}\label{calV1calV2}
\begin{align}
\label{calV(x)}
  & \hat{\mathcal{V}}(x)=\hat{\mathcal{V}}_1(x) + \hat{\mathcal{V}}_2(x) , \\
  \label{calV1(x)}
  & \hat{\mathcal{V}}_1(x)=\frac{Z_s^2}{1+\sigma^2}k^2\Delta\big[{\xi'}(x)\big]^2\ , \\
\label{calV2(x)}
  & \hat{\mathcal{V}}_2(x)= -i\frac{Z_s}{\sqrt{1+\sigma^2}}k
  \left[\xi'(x)\frac{\partial}{\partial x}+\frac{\partial}{\partial x}\xi'(x)\right]\ .
\end{align}
\end{subequations}
In view of the apparent similarity between equation \eqref{B(x)-eq} and one-dimensional Schr\"odinger equation, to obtain its solution we refer to the methods used earlier in the problems of Fermi particle and classical wave transport in strictly 1D random media ~\cite{Berez74, AbrikRyzh78, GorDorPrig83, KanerCheb87, FreiTar91, FreiTar01}.
In the present work, basing on methods used earlier in \cite{bib:Tar00,bib:MakTar01,TarShost15}, we apply a~further development of the technique.

The BCs for equation \eqref{B(x)-eq} follow from representations \eqref{H_0^lrc} and \eqref{in_trial field} for the trial function, namely,
\begin{subequations}\label{BCs_B(x)}
\begin{align}
\label{BC_B(L/2)}
 \begin{aligned}
  \begin{cases}
 & \mathcal{B}(L/2)=t_+\ ,\\
 & \mathcal{B}'(L/2)=ik_{\text{spp}}t_+\ ,
  \end{cases}
 \end{aligned}
\end{align}
\begin{align}
\label{BC_B(-L/2)}
\qquad\quad\
 \begin{aligned}
  \begin{cases}
 & \mathcal{B}(-L/2)=1+r_-\ ,\\
 & \mathcal{B}'(-L/2)=ik_{\text{spp}}(1-r_-)\ .
  \end{cases}
 \end{aligned}
\end{align}
\end{subequations}
We will only make use of conditions \eqref{BC_B(L/2)} given at ${x=L/2}$ since the conditions at $x=-L/2$ can be obtained through direct solution of dynamic equation \eqref{B(x)-eq}.

We will seek the solution to Eq.~\eqref{B(x)-eq} in the form of a sum of modulated exponentials ``running'' in opposite directions, viz.,
\begin{equation}\label{B(x)_WS}
  \mathcal{B}(x)=\pi(x)\mathrm{e}^{ik_{\text{spp}}x}
  -i\gamma(x)\mathrm{e}^{-ik_{\text{spp}}x}\ .
\end{equation}
Assuming the modulation factors $\pi(x)$ and $\gamma(x)$ to be smooth as compared to the adjacent exponentials (the so called weak scattering (WS) approximation), we arrive at a pair of coupled differential equations, viz.,
\begin{subequations}\label{pi_gamma-eqs}
\begin{align}
\label{pi_gamma-eq1}
 & \pi'(x)+ i\eta(x)\pi(x)+\zeta_-(x)\gamma(x)=0\ ,\\
\label{pi_gamma-eq2}
 & \gamma'(x)-i\eta(x)\gamma(x)+\zeta_+(x)\pi(x)=0\ .
\end{align}
\end{subequations}
These equations are of the first order, so we can set boundary conditions for them only at one of the ends of the interval $\mathbb{L}$, no matter at which one. Let us put the BC at the right end $x=L/2$. By matching functions \eqref{B(x)_WS} and \eqref{H_0^>} along with their derivatives at this point, we obtain
\begin{subequations}\label{BC_pi(+)gamma(+)}
\begin{align}
\label{BC_pi(+)}
  & \pi(L/2)=t_+\mathrm{e}^{-ik_{\text{spp}}L/2}\ ,\\[6pt]
\label{BC_gamma(+)}
  & \gamma(L/2)=0\ .
\end{align}
\end{subequations}
In a similar way, we obtain the BCs at the left end of $\mathbb{L}$,
\begin{subequations}\label{BC_pi(-)gamma(-)}
\begin{align}
\label{BC_pi(-)}
  & \pi(-L/2)=\mathrm{e}^{ik_{\text{spp}}L/2}\ ,\\
\label{BC_gamma(-)}
  & \gamma(-L/2)=ir_-\mathrm{e}^{-ik_{\text{spp}}L/2}\ .
\end{align}
\end{subequations}

Under WS approximation functions $\eta(x)$ and $\zeta_{\pm}(x)$ in \eqref{pi_gamma-eqs} can be represented as relatively narrow bunches of spatial harmonics of potential $\hat{\mathcal{ V}}(x)$, which are constructed using the ``smoothing'' procedure similar to the averaging method by Bogolyubov and Mitropolsky~\cite{Bogoliubov1961},
\begin{subequations}\label{eta_zeta-def}
\begin{align}
 \label{eta-def}
  & \eta(x) =\frac{1}{2 k_{\text{spp}}}\int\limits_{x-l}^{x+l} \frac{d t}{2 l} \mathrm{e}^{\mp ik_{\text{spp}}t}
  \hat{\mathcal{V}}(t)\mathrm{e}^{\pm ik_{\text{spp}} t}\ ,\\
 \label{zeta-def}
  & \zeta_{\pm}(x) =\frac{1}{2 k_{\text{spp}}}\int\limits_{x-l}^{x+l} \frac{d t}{2 l}
  \mathrm{e}^{\pm ik_{\text{spp}} t}
  \hat{\mathcal{V}}(t) \mathrm{e}^{\pm ik_{\text{spp}} t}\ .
\end{align}
\end{subequations}
Spatial interval $l$ in formulas \eqref{eta_zeta-def} has a size intermediate between the ``microscopic'' lengths of our problem, to which we refer the wavelength and the roughness correlation length, and a set of ``macroscopic'' characteristic lengths, which includes scattering and localization lengths, as well as the length of the rough segment $\mathbb{L}$,
\begin{equation}\label{l_interval}
  k^{-1},\,r_c\ll l \ll L^{(sc)},\,L^{(loc)},\,L\ .
\end{equation}
Note especially the importance of the position of the (identical) exponential factors in \eqref{zeta-def} with respect to potential $\hat{\mathcal{V}}(x)$ since $\mathrm{e}^{\pm ik_{\text{spp}} t}$ and $\hat{\mathcal{V}}(x)$ do not commute because of potential $\hat{\mathcal{V}}_2(x)$ containing the differential operators.
%
\subsection{Correlation properties of smoothed random potentials}
\label{Pots_corrs}
%
Statistical properties of randomly rough surfaces are normally specified in terms of the surface profile binary correlation function. As far as the essential functional variable in our problem is not that function itself, but rather its spatial derivative, we will define the basic correlation function by
\begin{equation}
  \AV{\xi'(x)\xi'(y)}=\sigma^2W(x-y)\ ,
\end{equation}
where function $W(x)\equiv W(|x|/2r_c)$ has unit maximal value at $x=0$ and decreases rapidly at a distance $|\Delta x| \sim r_c$. Strictly speaking, under conditions \eqref{l_interval} specific form of function $W(x)$ is not of crucial importance because this function only comes out in its integrated form, through correlators of effective ``potentials'' \eqref{eta_zeta-def}. Since these potentials in our problem serve as determinative statistical variables, it is desirable that correlation properties of the system be specified in terms of precisely these potentials.

Consider the following spatial correlator of random function $\eta(x)$,
\begin{equation}\label{<eta(x)eta(y)>}
  \mathcal{K}_{\eta\eta^*}(x,y)=\AV{\eta(x)\eta^*(y)}\ .
\end{equation}
Substituting the explicit form of $\eta(x)$ from \eqref{eta-def}, we get the equality
\begin{align}\label{K_etaeta*-corr}
  \mathcal{K}_{\eta\eta^*}(x,y)=\AV{\big[\eta_1(x)+\eta_2(x)\big]\big[\eta_1(y)+\eta_2(y)\big]^*}\ ,
\end{align}
where indexes 1 and 2 correspond to the pair of terms in \eqref{eta-def} appearing with representation \eqref{calV(x)}. If we assume (in order to simplify subsequent formulas) that random function $\xi(x)$ is distributed according to Gaussian law, then the only nonzero terms in \eqref{K_etaeta*-corr} take the form \vspace{\baselineskip}
\begin{widetext}
\begin{subequations}\label{<<eta1,eta2>>}
\begin{align}
\label{<eta1,eta1>}
  \Av{\eta_1(x)\eta_1^*(y)}=& \frac{k^4 |Z_s|^4}{2|k_{\text{spp}}|^2}\frac{\sigma^4}{(1+\sigma^2)^2}
  \int\limits_{x-l}^{x+l}\frac{dt_1}{2l}\int\limits_{y-l}^{y+l}\frac{dt_2}{2l}W^2(t_1-t_2)\ ,
\\
\label{<eta2,eta2>}
  \Av{\eta_2(x)\eta_2^*(y)}=& \frac{k^2 |Z_s|^2}{4|k_{\text{spp}}|^2} \frac{\sigma^2}{1+\sigma^2}
  \int\limits_{x-l}^{x+l}\frac{dt_1}{2l}\int\limits_{y-l}^{y+l}\frac{dt_2}{2l}W(t_1-t_2) \left[4|k_{\text{spp}}|^2 \mp 2i(k_{\text{spp}}+k^*_{\text{spp}})\frac{\partial}{\partial t_1}-
  \frac{\partial^2}{\partial t^2_1}\right]\ .
\end{align}
\end{subequations}
\end{widetext}
Taking into account that $r_c \ll l$, correlator \eqref{<eta1,eta1>} can be simplified to
\begin{equation}
\label{eta1,eta1-corr}
  \Av{\eta_1(x)\eta_1^*(y)} \approx \frac{1}{L_f^{(1)}}F_l(x-y)\approx \frac{1}{L_f^{(1)}}\delta(x-y)\ ,
\end{equation}
where function
\begin{equation}
  F_l(x)= \frac{1}{2l}\left(1-\frac{|x|}{2l}\right) \theta(2l-|x|)
\end{equation}
introduced earlier in \cite{bib:MakTar01} plays the role of a prelimit $\delta$-function on the scale of ``macroscopic'' lengths standing on the right-hand side of inequalities \eqref{l_interval}, and
\begin{equation}\label{Lf(1)}
  \frac{1}{L_f^{(1)}} = \frac{k^4 |Z_s|^4}{2|k_{\text{spp}}|^2} \frac{\sigma^4}{(1+\sigma^2)^2}
  \int\limits_{-\infty}^{\infty}dx W^2(x)
\end{equation}
is the reciprocal length of the ``forward'' scattering (corresponding to small one-fold transfer of the momentum) associated with potential $\hat{\mathcal{V}}_1(x)$~\cite{bib:MakTar98}. Correlator \eqref{<eta2,eta2>} can also be written in the form similar to \eqref{eta1,eta1-corr},
\begin{equation}\label{<eta2,eta2>-2}
  \Av{\eta_2(x)\eta_2^*(y)}
  \approx \frac{1}{L_f^{(2)}}F_l(x-y)\approx \frac{1}{L_f^{(2)}}\delta(x-y)\ ,
\end{equation}
where
\begin{equation}
\label{FS_length-2}
 \frac{1}{L_f^{(2)}} =
  k^2 |Z_s|^2\frac{\sigma^2}{1+\sigma^2}\int\limits_{-\infty}^{\infty}dx W(x)
\end{equation}
is the forward scattering rate associated with potential~$\hat{\mathcal{V}}_2(x)$.

The correlator
\begin{equation}\label{<zeta_pm(x)zeta_pm(y)>}
  \mathcal{K}_{\zeta_{\pm}\zeta_{\pm}^*}(x,y)=\Av{\zeta_{\pm}(x)\zeta_{\pm}^*(y)}\ ,
\end{equation}
analogously to the case of random field $\eta(x)$, is represented as a sum of two correlators corresponding to a~pair of terms in the r.h.s. of~\eqref{zeta-def},
\begin{widetext}
\begin{subequations}\label{<<zeta1,zeta2>>}
{\allowdisplaybreaks
\begin{align}
\label{<zeta1,zeta1>}
  \Av{\zeta_{\pm,1}(x)\zeta_{\pm,1}^*(y)}= & \frac{k^4 |Z_s|^4}{2|k_{\text{spp}}|^2} \frac{\sigma^4}{(1+\sigma^2)^2}
  \int\limits_{x-l}^{x+l}\frac{dt_1}{2l}\int\limits_{y-l}^{y+l}\frac{dt_2}{2l}
  \mathrm{e}^{\pm 2i(k_{\text{spp}}t_1-k^*_{\text{spp}}t_2)}W^2(t_1-t_2)\ , 
\\
\label{<zeta2,zeta2>}
  \Av{\zeta_{\pm,2}(x)\zeta_{\pm,2}^*(y)}=& \frac{k^2 |Z_s|^2}{4|k_{\text{spp}}|^2}\frac{\sigma^2}{1+\sigma^2}
  \int\limits_{x-l}^{x+l}\frac{dt_1}{2l}\int\limits_{y-l}^{y+l}\frac{dt_2}{2l}
  \mathrm{e}^{\pm 2i(k_{\text{spp}}t_1-k^*_{\text{spp}}t_2)} 
\notag\\
  & \phantom{\frac{k^2}{4|k_{\text{spp}}|^2}|Z_s|^2\frac{\sigma^2}{1+\sigma^2}
  \frac{k^2}{4|k_{\text{spp}}|^2}\ }
  \times\left[4|k_{\text{spp}}|^2 \mp2i(k_{\text{spp}}+k^*_{\text{spp}})\frac{\partial}{\partial t_1}-
  \frac{\partial^2}{\partial t^2_1}\right]W(t_1-t_2)\ .
\end{align}
}
\end{subequations}
Substituting correlation function $W(t_1-t_2)$ and its square in the form of Fourier integrals into \eqref{<<zeta1,zeta2>>}, we obtain
\vspace{\baselineskip}
\begin{subequations}
\begin{equation}
\label{<zeta1,zeta1>-2}
  \Av{\zeta_{\pm,1}(x)\zeta_{\pm,1}^*(y)} \approx
  \frac{1}{L_b^{(1)}}F_l(x-y)\ ,
\end{equation}
\vspace{-\baselineskip}
\begin{equation}
  \frac{1}{L_b^{(1)}} \approx   \frac{k^4 |Z_s|^4}{2|k_{\text{spp}}|^2}\frac{\sigma^4}{(1+\sigma^2)^2}
  \int\limits_{-\infty}^{\infty}\frac{dq}{2\pi}\widetilde{W}(q-k_{\text{spp}})\widetilde{W}^*(q+k_{\text{spp}}) \ ,
\end{equation}
\end{subequations}
\begin{equation}\label{<zeta2,zeta2>-2}
  \Av{\zeta_{\pm,2}(x)\zeta_{\pm,2}^*(y)}
  \approx \frac{k^2 |Z_s|^2}{4|k_{\text{spp}}|^2} \frac{\sigma^2}{1+\sigma^2}
  \widetilde{W}(\mp 2k_{\text{spp}})
  \int\limits_{-\infty}^{\infty}\frac{dq}{2\pi}
  \mathrm{e}^{i(q\pm 2k_{\text{spp}})(x-y)}
  \Big[4|k_{\text{spp}}|^2\pm 4qk'_{\text{spp}}+q^2\Big]
  \frac{\sin^2[(q\pm 2k_{\text{spp}})l]}{[(q\pm 2k_{\text{spp}})l]^2}\ .
\end{equation}
\end{widetext}
The last factor under the integral in \eqref{<zeta2,zeta2>-2} is a ``sharp'' function with maxima at points ${q_m=\mp 2k_{\text{spp}}}$. If one moves the factor in square brackets outside the integral at any of these two points, then the result of calculating the integral will become zero. This implies that the ``backward'' scattering corresponding to twice the momentum $k_{\text{spp}}$ transfered in each scattering event, in the WS limit is caused by potential $\hat{\mathcal{V}}_1(x)$ only. The backward scattering length due to this potential, which is usually associated with one-dimensional localization length, is given by
\begin{equation}\label{Lb->L_loc}
  \frac{1}{L_b} \approx
  \frac{k^4 |Z_s|^4}{2|k_{\text{spp}}|^2}\frac{\sigma^4}{(1+\sigma^2)^2}
  \int\limits_{-\infty}^{\infty}dx \mathrm{e}^{-2ik_{\text{spp}}x}W^2(x)\ .
\end{equation}
%
\subsection{Some preliminary estimates}
\label{Quant_estims}
%
Since for a good metal the inequality $|Z_s|\lesssim 1$ is usually fulfilled \cite{Palik98}, we can evaluate the scattering length from Eq.~\eqref{Lf(1)} as
\begin{subequations}\label{Scatt_lengths}
\begin{equation}\label{L1(f)-est}
  \frac{1}{L_f^{(1)}}\sim k^2 r_c\frac{\sigma^4}{1+\sigma^4}\ .
\end{equation}
The estimate for length $L_f^{(2)}$ is not very different from \eqref{L1(f)-est}, namely,
\begin{equation}\label{L2(f)-est}
  \frac{1}{L_f^{(2)}}\sim k^2 r_c\frac{\sigma^2}{1+\sigma^2}\ .
\end{equation}

The estimate for the ``backward'' scattering length $L_b$ depends essentially on the value of parameter $kr_c$. For $kr_c\ll 1$ it coincides with \eqref{L1(f)-est}. In the opposite limiting case, $kr_c\gg 1$, the oscillating exponential under the integral in \eqref{Lb->L_loc} leads to its significant decrease, and evaluation is as follows,
\begin{equation}\label{L1(b)-est}
  \frac{1}{L_b}\sim \frac{1}{L_f^{(1)}} \cdot\frac{1}{1+kr_c}\ .
\end{equation}
\end{subequations}
In our calculations, approximation \eqref{l_interval} plays a significant role, implying that all scattering lengths are large as compared to the wavelength. From this fact it follows that
\begin{equation}\label{WS_lengths}
  k r_c\frac{\sigma^2}{1+\sigma^2} \ll 1\ ,
\end{equation}
where parameter $\sigma^2$ characterizes the degree of mean sharpness of surface asperities. If the roughness is on average smooth ($\sigma^2\ll 1$), condition \eqref{WS_lengths} is reduced to inequality
\begin{equation}\label{Large_Frenel}
  \frac{k h^2}{r_c}\ll 1\ ,
\end{equation}
where $h$ is the characteristic height of the asperities. Inequality \eqref{Large_Frenel}, in the language of wave optics, corresponds to large values of Fresnel parameter, i.\,e., to the Fraunhofer diffraction which does not take into account the local curvature of the scattering surface. In the opposite limit, where the asperities are strongly sharpened ($\sigma^2\gg 1$), all the results obtained below are valid under condition that $kr_c\ll 1$.
%
\section{On the scattering amplitude asymptotics}
\label{ScatAmpl-asymp}
%
Although formula \eqref{FredholmEq-oper} for the scattering amplitude is quite complicated to operate with it in the general form, its algebraic structure allows the interpretation in terms of either single or multiple scattering by fluctuations of the effective impedance through some intermediate states we entitle \textit{composite plasmons} (CP) \cite{TarStadKvit23}. The meaning of this term in our study is related to the operator $\hat{\mathcal{L}}$ representation in \eqref{FredholmEq-oper} in terms of a product of two operators, $\hat{G}^{\scriptscriptstyle {(\text{CP})}}$ and $\Delta\hat{\overline{Z}}_s$, whose matrix elements in momentum representation are given by
\begin{subequations}\label{hatG(CP)*hat_Delta_Zs}
\begin{align}\label{hatG(CP)}
 & \bra{q}\hat{G}^{\scriptscriptstyle{(\text{CP})}}\ket{q'} =
  Z_{e}^{-1}\left( q\right) 2\pi\delta(q-q')\ ,\\
\label{hat_Delta_Zs}
  & \bra{q}\Delta\hat{\overline{Z}}_s\ket{q'} =\Delta\widetilde{\overline{Z}}_s(q-q')\ .
\end{align}
\end{subequations}
Operator $\hat{G}^{\scriptscriptstyle{(\text{CP})}}$ is a propagator of wave excitations representing the weighted superposition of the SPP-type surface wave and a bunch of bulk harmonics propagating quasi-isotropically in free space above the metal surface. The latter harmonics can readily be associated with outward-scattered leaky waves. Operator $\Delta\hat{\overline{Z}}_s$ is related to the impedance non-uniformity, which causes the SPP and volume harmonics to couple to each other.

Such a structure of operator $\hat{\mathcal{L}}$ gives grounds to consider it as the \textit{mixing} operator for different one-dimensional harmonics of the trial field, which are of purely surface nature, and the outward-radiated volume harmonics having nonzero real $z$-component of the wave vector. The degree of coupling of surface and bulk harmonics in each single scattering act can be characterized by the norm (analog of value) of the operator $\hat{\mathcal{L}}$. We evaluate this norm in the next section.
%
\subsection{Evaluation of the operator $\bm{\hat{\mathcal{L}}}$ norm}
\label{L_oper_norm}
%
We will characterize the random operator $\hat{\mathcal{L}}$ ``magnitude'' by its average square Euclidean norm \cite{Rudin1991}
\begin{equation}\label{Norm-def}
  \Av{\|\hat{\mathcal{L}}\|^2}=\sup_{\varphi}
  \frac{\Av{\big(\hat{\mathcal{L}}\varphi,\hat{\mathcal{L}}\varphi\big)}}{\big(\varphi,\varphi\big)}\ .
\end{equation}
Here $\varphi$ is a complete set of functions on which the action of operator $\hat{\mathcal{L}}$ is defined. Order-of-magnitude estimate of \eqref{Norm-def} can be obtained by comparing the results of action of operator $\hat{\mathcal{L}}=\hat{G}^{\scriptscriptstyle{(\text{CP})}}\Delta\hat{\overline{Z}}_s$ and the identity operator on the same arbitrary vector $\varphi(x)$ of some Banach space $\mathbb{B}\{\varphi\}$ whose properties we are going to specify. It is convenient to work with operator $\hat{\mathcal{L}}$ in $q$-representation, in which its matrix elements are given by~\eqref{Kernel_Fredholm_eq}.

The operator $\hat{\mathcal{L}}$ action on an arbitrary function ${\widetilde{\varphi}(q)\in\mathbb{B}}$, in view of finite support of function $\Delta\overline{Z}_s( x)$, is defined by equality
\begin{equation}\label{L->phi(x)}
  \big[\hat{\mathcal{L}}\varphi\big](q)=
  \int\limits_{-\infty}^{\infty}\frac{dq'}{2\pi}
  \mathcal{L}(q,q')\widetilde{\varphi}(q')
  =\int_{\mathbb{L}}dx\mathcal{L}(q,x)\varphi(x)\ .
\end{equation}
We will be interested in the class of functions decomposable into Fourier integrals on segment $\mathbb{L}$, but not necessarily possessing fixed values at the ends of this segment. The last condition is imposed in order to permit consideration of boundary segments with open ends.

\begin{figure}[h!!!]
  \centering
  \scalebox{.8}[.8]{\includegraphics{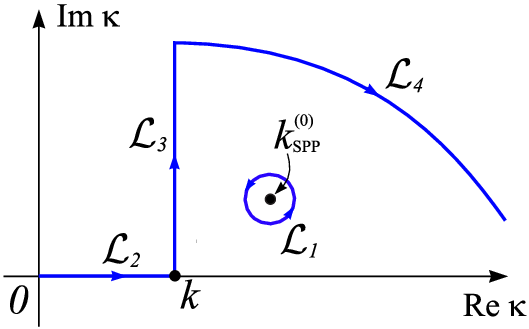}}
  \caption{Deformation of the integration contour in \eqref{Int_kappa_est}.}
  \hfill\label{fig2}
\end{figure}

The action of an arbitrary operator $\hat{\mathcal{A}}$ on function $\varphi(x)$ from the aforementioned class is given by equality
\vspace{2\baselineskip}
\begin{widetext}
\begin{align}\label{hat_A->phi}
  \big(\hat{\mathcal{A}}\varphi\big)(q)=& \int_{\mathbb{L}}dx'\bra{q}\hat{\mathcal{A}}\ket{x'}\varphi(x')=
  \int\limits_{-\infty}^{\infty}\frac{d\kappa}{2\pi}\widetilde{\varphi}(\kappa)
  \int_{\mathbb{L}}dx'\bra{q}\hat{\mathcal{A}}\ket{x'}\mathrm{e}^{i\kappa x'}\ .
\end{align}
The norm squared of vector $\hat{\mathcal{A}}\varphi$, defined through the scalar product, is
\begin{align}\label{|A|^2}
  \|\hat{\mathcal{A}}\varphi\|^2 &=\big(\hat{\mathcal{A}}\varphi,\hat{\mathcal{A}}\varphi\big)=
  \iint\limits_{-\infty}^{\quad\infty}\frac{d\kappa_1d\kappa_2}{(2\pi)^2}
  \widetilde{\varphi}^*(\kappa_1)\widetilde{\varphi}(\kappa_2)
  \int\limits_{-\infty}^{\infty}\frac{dq}{2\pi}\iint_{\mathbb{L}}dx_1dx_2
  \bra{q}\hat{\mathcal{A}}\ket{x_1}^*\bra{q}\hat{\mathcal{A}}\ket{x_2}
  \mathrm{e}^{-i\kappa_1x_1+i\kappa_2x_2}\ .
\end{align}
%
If $\hat{\mathcal{A}}$ be identity operator, then
\begin{align}\label{|1|^2}
  \|\hat{\mathbf{1}}\cdot\varphi\|^2=
  \iint\limits_{-\infty}^{\quad\infty}\frac{d\kappa_1d\kappa_2}{(2\pi)^2}
  \widetilde{\varphi}^*(\kappa_1)\widetilde{\varphi}(\kappa_2)\Delta_L(\kappa_1-\kappa_2)\ , 
\end{align}
where
\begin{align}\label{Underlimit_delta}
  \Delta_L(\kappa)=\int\limits_{-L/2}^{L/2}dx\,\mathrm{e}^{\pm i\kappa x}
  =\frac{\sin(\kappa L/2)}{\kappa/2}
\end{align}
is the function which plays the role of pre-limit $2\pi\delta$-function having characteristic width $\sim 1/L$,
\begin{equation*}
  \Delta_L(\kappa)  \xrightarrow[L\to\infty]{}2\pi\delta(\kappa)
  \quad\text{when}\quad \mathrm{Im}\,\kappa= 0.
\end{equation*}
If in place of the operator $\hat{\mathcal{A}}$ we take operator $\hat{\mathcal{L}}$ with matrix elements \eqref{Kernel_Fredholm_eq}, then its action on the oscillating exponent supported by $\mathbb{L}$ is given by expression
\begin{align}\label{^L->e^(iqx)}
  \big[\hat{\mathcal{L}}\cdot\mathrm{e}^{i\kappa_j x'}\big](\kappa) &=
  \int\limits_{-\infty}^{\infty}\frac{d\kappa'}{2\pi}
  \bra{\kappa}\hat{G}^{\scriptscriptstyle{(\text{CP})}}\ket{\kappa'}\int_{\mathbb{L}}dx'\bra{\kappa'}
  \Delta\hat{\overline{Z}}_s\ket{x'}
  \mathrm{e}^{i\kappa_jx'}=
  Z_{e}^{-1}\left( \kappa\right) \Delta\widetilde{\overline{Z}}_s(\kappa-\kappa_j)\ .
\end{align}
The norm squared of vector $\hat{\mathcal{L}}\varphi$ then becomes
\begin{align}\label{|L|^2}
  \|\hat{\mathcal{L}}\varphi\|^2=& \iint\limits_{-\infty}^{\quad\infty}\frac{d\kappa_1d\kappa_2}{(2\pi)^2}
  \widetilde{\varphi}^*(\kappa_1)\widetilde{\varphi}(\kappa_2)
  \int\limits_{-\infty}^{\infty}\frac{d\kappa}{2\pi}
  \frac{\Delta\widetilde{\overline{Z}}^*_s(\kappa-\kappa_1)\Delta\widetilde{\overline{Z}}_s(\kappa-\kappa_2)}{\left|Z_{e}\left( \kappa\right)\right|^{2}} =
\notag\\
  =& \iint\limits_{-\infty}^{\quad\infty}\frac{d\kappa_1d\kappa_2}{(2\pi)^2}
  \widetilde{\varphi}^*(\kappa_1)\widetilde{\varphi}(\kappa_2)  \iint_{\mathbb{L}}dx_1dx_2\mathrm{e}^{-i\kappa_1x_1+i\kappa_2x_2}
  \Delta\overline{Z}^*_s(x_1)\Delta\overline{Z}_s(x_2)
  \int\limits_{-\infty}^{\infty}\frac{d\kappa}{2\pi}
  \frac{\mathrm{e}^{i\kappa(x_1-x_2)}}{\left|Z_{e}\left( \kappa\right)\right|^{2}}\ .
\end{align}
\end{widetext}
To average this square norm we have to perform the averaging of rather complicated functional of random function $\xi'(x)$. During this procedure, a correlation function arises which can be with high accuracy considered as being dependent on the \textit{difference} of arguments $x_{1}$ and~$x_{2}$. By denoting
\begin{equation}\label{Q(x1-x2)}
  \mathcal{Q}(x_1-x_2)= \Av{\Delta\overline{Z}^*_s(x_1)\Delta\overline{Z}_s(x_2)}\ ,
\end{equation}
we obtain
\begin{widetext}
\begin{align}\label{<|L|^2>}
  \Av{\|\hat{\mathcal{L}}\varphi\|^2}\approx
  \iint\limits_{-\infty}^{\quad\infty}\frac{d\kappa_1d\kappa_2}{(2\pi)^2}
  \widetilde{\varphi}^*(\kappa_1)\widetilde{\varphi}(\kappa_2)\Delta_L(\kappa_1-\kappa_2)
  \int\limits_{-L}^L dx \mathcal{Q}(x)\mathrm{e}^{-i\kappa_1 x}
  \int\limits_{-\infty}^{\infty}\frac{d\kappa}{2\pi}
  \frac{\mathrm{e}^{i\kappa x}}{\left|Z_{e}\left( \kappa\right)\right|^{2}}\ .
\end{align}
\end{widetext}
Correlation function $\mathcal{Q}(x)$ decays on a scale of the order of scattering length, $\Delta_{\scriptscriptstyle{\mathcal{Q}}}x\sim L^{(sc)}$, which under weak scattering is much larger than the wavelength. To estimate the scale of the change in variable $x$ of the last, integral factor in \eqref{<|L|^2>} it is necessary to calculate, or at least to estimate by the order of magnitude, the integral over $\kappa$. This is convenient to do by presenting it in the form
\begin{equation}\label{Int_kappa_est}
  \mathcal{I}(x) = \Re \int\limits_{0}^{\infty}\frac{d\kappa}{\pi}
  \frac{\mathrm{e}^{i\kappa |x|}}{\left|Z_{e}\left( \kappa\right)\right|^{2}}
\end{equation}
and applying the methods of the theory of complex variables. In the first quadrant of complex~$\kappa$, the integrand of \eqref{Int_kappa_est} has two exceptional points. The first is a pole at point
\begin{equation}\label{k_spp(0)}
  k_{\text{spp}}^{\text{(0)}}=k\sqrt{1-\left(\overline{Z}_s^{(0)}\right)^2}\ ,
\end{equation}
which differs from the unperturbed wave number $k_{\text{spp}}$ from \eqref{RightLeft_SPP} by renormalization of the effective impedance entering into it (see formula \eqref{Zs(0)}). The second is a branch point at $\kappa=k$, which is located directly on the integration axis. By deforming the integration contour as shown in Fig.~\ref{fig2},
we reduce the problem of estimating the required square norm to estimating the value of integrals over four contours denoted in the figure by symbols $\mathcal{L}_i$, $i=1,2,3,4$.

The result of integration over contour $\mathcal{L}_1$ is determined by the residue at point $k_{\text{spp}}^{\text{\tiny (0)}}$ and is equal to
%
\begin{align}\label{I(1)}
  \mathcal{I}_1(x)=&  k \frac{{Z}'_s}{\left|{Z}''_s\right|} \cdot
  \frac{\sqrt{1+\Big(\Big[\overline{Z}_s^{(0)}\Big]''\Big)^2}+\Big[\overline{Z}_s^{(0)}\Big]''}
  {1+\Big(\Big[\overline{Z}_s^{(0)}\Big]''\Big)^2}  \mathrm{e}^{ik_{\text{spp}}^{\text{\tiny (0)}}|x|}
\notag\\
 & \sim k \frac{{Z}'_s}{\left|{Z}''_s\right|} \mathrm{e}^{ik_{\text{spp}}^{\text{\tiny (0)}}|x|}\ .
\end{align}
%
The integral over contour $\mathcal{L}_2$ is
\begin{equation}\label{I_2-int}
  \mathcal{I}_2(x)= \int\limits_{0}^{k}\frac{d\kappa}{\pi}
  \frac{\cos{\kappa x}}{\left|Z_{e}\left( \kappa\right)\right|^{2}}=
  \Re \int\limits_{0}^{k}\frac{d\kappa}{\pi}
  \frac{\mathrm{e}^{i\kappa |x|}}{\left|Z_{e}\left( \kappa\right)\right|^{2}}\ .
\end{equation}
The characteristic values of coordinate $x$ here are established by the integrand in \eqref{<|L|^2>}. The first value is the length of the rough boundary segment, $L$. The second one is the localization length $L^{(sc)}$ characteristic of correlator  $\mathcal{Q}(x)$. And the third value is the localization scale of trial functions $\widetilde{\varphi}(\kappa)$, which is of the same order of magnitude as~$L^{(sc)}$. As a result, the oscillation period of the integrand in \eqref{I_2-int} turns out to be small compared to the integration interval, and the order-of-magnitude estimate of the integral is as follows:
\begin{equation}\label{I_2-estim}
  \mathcal{I}_2(x)\sim \frac{1}{L^{\text{(macr)}}}\ .
\end{equation}
Here, by $L^{\text{(macr)}}$ we mean the smallest of the ``macroscopic'' lengths of our problem, viz.,
\begin{equation}\label{L(macr)}
  L^{\text{(macr)}}= \min \left(L, L^{(sc)}, L^{\scriptscriptstyle{(\text{SPP})}}_{dis}\right)\ .
\end{equation}
The integral over contour $\mathcal{L}_3$ differs from integral \eqref{I_2-int} by the integration limits,
\begin{equation}\label{I_3-int}
  \mathcal{I}_3(x)=
  \Re \int\limits_{k}^{k+i\infty}\frac{d\kappa}{\pi}
  \frac{\mathrm{e}^{i\kappa |x|}}{\left|Z_{e}\left( \kappa\right)\right|^{2}}\ ,
\end{equation}
so its estimate is similar to \eqref{I_2-estim}. The integral over contour $\mathcal{L}_4$, with account of its (infinite) distance from the coordinate origin, can be set approximately equal to zero.

Given the definition of the  SPP dissipative damping length \eqref{L(SPP)_dis}, the estimate \eqref{I_2-estim} can be considered common for all fragments of integral \eqref{Int_kappa_est} and hence for the integral as a whole. Substituting this estimate into \eqref{<|L|^2>} and taking account of \eqref{|1|^2}, we obtain the following estimate for operator~$\hat{\mathcal{L}}$ mean squared norm,
\begin{equation}\label{Norm_L-prefin}
  \Av{\|\hat{\mathcal{L}}\|^2} \sim \Big<{\left|\Delta\overline{Z}_s(x)\right|^2}\Big>\ .
\end{equation}
With the definition of quantity $\Delta\overline{Z}_s(x)$ and taking account of \eqref{Z^(eff)}, we arrive at the following estimation,
\begin{equation}\label{Norm_L-fin}
  \Av{\|\hat{\mathcal{L}}\|^2} \sim \frac{\left|{Z}_s\right|^2}{1+\sigma^2}\ .
\end{equation}
Since in metals the inequality normally holds true ${\left|{Z}_s\right|\lesssim 1}$, it is quite acceptable to use the inequality ${\Av{\|\hat{\mathcal{L}}\|}\ll 1}$ when calculating the radiation field analytically. Note that the more ``sharp'' (on average) is the degree of roughness of the metal boundary, the more this approximation is justified.
%
\section{The scattering pattern for an incident surface plasmon polariton}
%
Estimate \eqref{Norm_L-fin} allows to examine with good accuracy the SPP scattering by a~randomly corrugated segment of metal boundary in the approximation of weak intermixing of surface and bulk scattered modes, which corresponds to inequality
\begin{equation}\label{Weak-Mix_cond}
  \Av{\|\hat{\mathcal{L}}\|}\ll 1\ .
\end{equation}
The average intensity of leaking field \eqref{Full_solution_h(r)} is given in general by expression
\begin{subequations}
\begin{equation}\label{|h(r)|^2-gen}
  \Av{|h(\mathbf{r})|^2}=
  \frac{1}{2}\iint\limits_{-\infty}^{\quad\infty}\frac{dqdq'}{(2\pi)^2}
  \Psi(q,q',x,z) \mathcal{R}(q,q') ,
\end{equation}
where
\begin{align}
 \label{Psi(q,q')}
 & \Psi(q,q',x,z)= e^{i(q-q')x+i\left[\sqrt{k^2-q^2}-\left(\sqrt{k^2-q'^2}\right)^*\right]z} ,\\[6pt]
 \label{R(q,q')}
 & \mathcal{R}(q,q') = \Av{\widetilde{\mathcal{R}}(q)\widetilde{\mathcal{R}}^*(q')}.
\end{align}
\end{subequations}
In the approximation \eqref{Weak-Mix_cond} from \eqref{FredholmSol-momentum} we have
\begin{equation}\label{<R(q)R(q')>}
  \mathcal{R}(q,q')
  \approx \iint\limits_{-\infty}^{\quad\infty}\frac{dq_1dq_2}{(2\pi)^2}
  \left<\mathcal{L}(q,q_1)\mathcal{L}^*(q',q_2)
  \widetilde{\mathcal{B}}(q_1)\left.\widetilde{\mathcal{B}}\right.^*(q_2)\right>.
\end{equation}
This formula can be simplified if we notice that the random functions in operator $\hat{\mathcal{L}}$ matrix elements are fundamentally different from functions $\widetilde{\mathcal{B}}(q_{1,2})$.
Factor $\widetilde{\mathcal{B}}(q)$ is a~functional of effective spatially integrated (``smoothed'') random potentials \eqref{eta_zeta-def}, while random impedance $\Delta\widetilde{\overline{Z}}_s(q)$ enters into operator $\hat{\mathcal{L }}$ kernel in a pure, unintegrated form. The spatial scale of the substantial variation of function $\Delta\overline{Z}_s(x)$, as it can be seen from definition \eqref{Z^(eff)}, is either macroscopic (for $\sigma\ll 1$ this function with good accuracy can be considered constant), or (for $\sigma\gtrsim 1$) has an order of magnitude of the roughness correlation length.

When coupling (in the course of averaging) this function with smoothed potentials $\eta(x)$ and $\zeta_{\pm}(x)$, on which factor~$\mathcal{B}(x)$ depends functionally, the effective $\delta$-functions similar to those appearing in the correlation relations for $\eta(x)$ and $\zeta_{\pm}(x)$ (see section \ref{Pots_corrs}) are not formed. As a~result, such a~pairing is subjected to an additional spatial averaging which, under conditions \eqref{l_interval}, leads to its excessive reduction by one of the two small factors: either $1/kl\ll 1$ or $\sim r_c/l\ll 1 $. Therefore, the correlator in \eqref{<R(q)R(q')>} can be with asymptotic accuracy factorized into correlators $\bm{\mathcal{K}}_1$ and $\bm{\mathcal {K}}_2$ of the following form,
\begin{subequations}\label{K_1K_2-corrs}
\begin{align}
\label{K_1-corr}
  & \bm{\mathcal{K}}_1(q,q'|q_1,q_2)= \Big<\mathcal{L}(q,q_1)\mathcal{L}^*(q',q_2)\Big>\ ,
\\
\label{K_2-corr}
  & \bm{\mathcal{K}}_2(q_1,q_2)= \left<\widetilde{\mathcal{B}}(q_1)\left.\widetilde{\mathcal{B}}\right.^*(q_2)\right>\ .
\end{align}
\end{subequations}
The calculation of these correlators is given in Appendix~\ref{K_1K_2-calc}, and the integral over $q_1$ and $q_2$ of their product is represented by formula \eqref{Int_q1q2_K1K2-3}. As a result, we arrive at the following expression for average intensity~\eqref{|h(r)|^2-gen},
\vspace{-\lineskip}
\begin{widetext}
{\allowdisplaybreaks
\begin{align}\label{|h(r)|^2-gen(2)}
  \Av{|h(\mathbf{r})|^2}\approx & \frac{\sigma^4}{4(1+\sigma^2)^2}\left|\overline{Z}_s^{(0)}\right|^2 \iint\limits_{-\infty}^{\quad\infty}\frac{dqdq'}{(2\pi)^2}
  \frac{\Psi(q,q',x,z)}
  {Z_{e}(q)Z_{e} ^*(q')}
\notag\\[3pt]
 & \times \int_{\mathbb{L}}d\xi
  \bigg\{\Av{|\pi(\xi)|^2}\big[\widetilde{W}^2\big](q-k_{\text{spp}})+
  \Av{|\gamma(\xi)|^2}\big[\widetilde{W}^2\big](q+k_{\text{spp}}) 
\notag\\[6pt]
 & \qquad + i\Av{\pi(\xi)\gamma^*(\xi)}\big[\widetilde{W}^2\big](q-k_{\text{spp}})-
  i\Av{\gamma(\xi)\pi^*(\xi)}\big[\widetilde{W}^2\big](q+k_{\text{spp}})\bigg\}
  \mathrm{e}^{-i(q-q')\xi}\ .
\end{align}
}
\end{widetext}
%
\subsection{Calculation of the one-point correlators of quasi-amplitudes $\bm{\pi(x)}$ and $\bm{\gamma(x)}$}
\label{ScattDiagr}
%
To average random functions $|\pi(x)|^2$, $|\gamma(x)|^2$, $\pi(x)\gamma^*(x)$, and $\gamma(x) \pi^*(x)$ entering into \eqref{|h(r)|^2-gen(2)} we make use of equations \eqref{pi_gamma-eqs}. Neglecting the dissipation in the metal, potentials \eqref{calV1calV2} can be considered real ($Z_s$ in this case is a~purely imaginary quantity). Random field $\eta(x)$ is then real as well, while fields $\zeta_{\pm}(x)$ are complex conjugate to each other.

Introducing new random functions $\widetilde{\pi}(x)$ and $\widetilde{\gamma}(x)$ through phase relationships
\begin{subequations}\label{-pi-gamma}
{\allowdisplaybreaks
\begin{align}\label{-pi}
 & \widetilde{\pi}(x)=\pi(x)\exp\Bigg[-i\int\limits_x^{L/2}dx'\eta(x')\Bigg]\ , \\[3pt]
\label{-gamma}
 & \widetilde{\gamma}(x)=\gamma(x)\exp\Bigg[i\int\limits_x^{L/2}dx'\eta(x')\Bigg]\ ,
\end{align}
}
\end{subequations}
we obtain a pair of coupled equations
\vspace{.7\baselineskip}
\begin{subequations}\label{-pi-gamma-eqs}
\begin{align}
\label{pi_gamma-eq1(2)}
 & \widetilde{\pi}'(x)+\widetilde{\zeta}_-(x)\widetilde{\gamma}(x)=0\ ,\\[6pt]
\label{pi_gamma-eq2(2)}
 & \widetilde{\gamma}'(x)+\widetilde{\zeta}_+(x)\widetilde{\pi}(x)=0\ ,
\end{align}
\end{subequations}
where the renormalized random fields $\widetilde{\zeta}_{\pm}(x)$ are related to the original fields $\zeta_{\pm}(x)$ by equalities
\begin{equation}\label{-zeta_pm}
  \widetilde{\zeta}_{\pm}(x)=\zeta_{\pm}(x)\exp\Bigg[\pm 2i\int\limits_x^{L/2}dx'\eta(x')\Bigg]\ .
\end{equation}
Due to effective $\delta$-correlation of the original random fields $\zeta_{\pm}(x)$ and $\zeta^*_{\pm}(x)$, see Eq.~\eqref{<zeta1,zeta1>-2}, correlation properties of these new fields do not change. Yet, equations \eqref{-pi-gamma-eqs} contain in the explicit form a smaller number of ``independent'' random functions than equations \eqref{pi_gamma-eqs}, and therefore it will be more convenient to operate further just with Eqs.~\eqref{-pi-gamma-eqs}.

With notations \eqref{-pi-gamma}, formula \eqref{|h(r)|^2-gen(2)} for the leaking field intensity takes the form
\begin{widetext}
\begin{align}\label{|h(r)|^2-gen(wav)}
  \Av{|h(\mathbf{r})|^2}\approx & \frac{\sigma^4}{4(1+\sigma^2)^2}\left|\overline{Z}_s^{(0)}\right|^2 \iint\limits_{-\infty}^{\quad\infty}\frac{dqdq'}{(2\pi)^2} \frac{\Psi(q,q',x,z)} {Z_{e}(q)Z_{e} ^*(q')}
\notag\\
 & \times \int_{\mathbb{L}}d\xi
  \Bigg\{\AV{|\widetilde{\pi}(\xi)|^2}\big[\widetilde{W}^2\big](q-k_{\text{spp}})+
  \AV{|\widetilde{\gamma}(\xi)|^2}\big[\widetilde{W}^2\big](q+k_{\text{spp}})
\notag \\
 &\phantom{\int_{\mathbb{L}}d\xi\bigg\{\quad }
  \ + i\AV{\widetilde{\pi}(\xi)\widetilde{\gamma}^*(\xi)
  \exp\bigg[2i\int_{\xi}^{L/2}dx'\eta(x')\bigg] } \big[\widetilde{W}^2\big](q-k_{\text{spp}})
\notag \\
 & \phantom{\int_{\mathbb{L}}d\xi\bigg\{\quad }
 \ - i\AV{\widetilde{\gamma}(\xi)\widetilde{\pi}^*(\xi)
 \exp\bigg[-2i\int_{\xi}^{L/2}dx'\eta(x')\bigg] } \big[\widetilde{W}^2\big](q+k_{\text{spp}})\Bigg\}
  \mathrm{e}^{-i(q-q')\xi}\ .
\end{align}
\end{widetext}
For averaging the product $\widetilde{\pi}(\xi)\widetilde{\gamma}^*(\xi)$ and its conjugate counterpart with oscillating exponents in~\eqref{|h(r)|^2-gen(wav)}, one should take into account the statistical independence of random fields $\widetilde{\zeta} _{\pm}(x)$, on the one hand, and random field $\eta(x)$ on the other. This assumes that for the last two terms in the integral over~$\xi$ in \eqref{|h(r)|^2-gen(wav)} the averaging of the exponentials and pre-exponential factors $\widetilde{\pi}(\xi)\widetilde{\gamma}^*(\xi)$ and $\widetilde{\gamma}(\xi)\widetilde{\pi}^*(\xi)$ can be performed independently. Under conditions \eqref{l_interval}, the averaging of the oscillatory exponentials is carried out rigorously, resulting with asymptotic accuracy in
\begin{align}\label{Forward_scat_exp}
  \AV{\exp\bigg[2i\int_{\xi}^{L/2}dx'\eta(x')\bigg] }=
  \exp\bigg[-\frac{2}{L_f}\big(L/2-\xi\big)\bigg]\ .
\end{align}
As for the rest of random functions in \eqref{|h(r)|^2-gen(wav)}, we will describe the method of their averaging in more details.

From Eqs. \eqref{-pi-gamma-eqs} and \eqref{-pi-gamma} we obtain the following set of differential equations,
\begin{subequations}\label{|P|^2,|G|^2,PG-eqs}
\begin{align}
\label{|P|^2--eq}
 & \frac{d\,|\widetilde{\pi}(x)|^2}{dx}=-\widetilde{\zeta}_-(x)\widetilde{\gamma}(x)\widetilde{\pi}^*(x)-
  \widetilde{\zeta}_+(x)\widetilde{\pi}(x)\widetilde{\gamma}^*(x)\ ,\\
\label{|G|^2--eq}
 & \frac{d\,|\widetilde{\gamma}(x)|^2}{dx}=-\widetilde{\zeta}_-(x)\widetilde{\gamma}(x)\widetilde{\pi}^*(x)-
  \widetilde{\zeta}_+(x)\widetilde{\pi}(x)\widetilde{\gamma}^*(x)\ ,\\
\label{PG*--eq}
 & \frac{d}{dx}\Big[\widetilde{\pi}(x)\widetilde{\gamma}^*(x)\Big]=
 -\widetilde{\zeta}_-(x)\Big[|\widetilde{\pi}(x)|^2+|\widetilde{\gamma}(x)|^2\Big]\ ,\\
\label{GP*--eq}
 & \frac{d}{dx}\Big[\widetilde{\gamma}(x)\widetilde{\pi}^*(x)\Big]=
 -\widetilde{\zeta}_+(x)\Big[|\widetilde{\pi}(x)|^2+|\widetilde{\gamma}(x)|^2\Big]\ .
\end{align}
\end{subequations}
After averaging them, the correlators which appear on their right-hand sides can be transformed, applying Furutsu-Novikov method \cite{Furutsu63,Novikov64}, to the following form,
\begin{subequations}\label{<ze_pm_ga_pi*>,<ze_pm|pi|^2>,<ze_pm|ga|^2>}
{\allowdisplaybreaks
\begin{align}
\label{<ze_pm_ga_pi*>}
 & \left<\widetilde{\zeta}_-(x)\widetilde{\gamma}(x)\widetilde{\pi}^*(x)\right>=
  \left<\widetilde{\zeta}_+(x)\widetilde{\pi}(x)\widetilde{\gamma}^*(x)\right>=
\notag\\
 & \phantom{\widetilde{\zeta}_-(x)\widetilde{\gamma}(x) \frac{1}{2L_b}}
  = \frac{1}{2L_b}\Big\{\left<|\widetilde{\pi}(x)|^2\right>+\left<|\widetilde{\gamma}(x)|^2\right>\Big\}\ ,\displaybreak[0]\\[10pt]
\label{<ze_pm|pi|^2>}
 & \AV{\widetilde{\zeta}_-(x)\Big[|\widetilde{\pi}(x)|^2+|\widetilde{\gamma}(x)|^2\Big]}=
 \frac{1}{L_b}\AV{\Big[\widetilde{\pi}(x)\widetilde{\gamma}^*(x)\Big]}\ ,\\
\label{<ze_pm|ga|^2>}
 & \AV{\widetilde{\zeta}_+(x)\Big[|\widetilde{\pi}(x)|^2+|\widetilde{\gamma}(x)|^2\Big]}=
 \frac{1}{L_b}\AV{\Big[\widetilde{\gamma}(x)\widetilde{\pi}^*(x)\Big]}\ .
\end{align}
}
\end{subequations}
Using these relationships we arrive at a set of equations which are pairwise closed with respect to the desired averages, viz.,
{
\allowdisplaybreaks
\begin{subequations}\label{<|P|^2>,<|G|^2>,<PG>-fin}
\begin{align}
\label{<|P|^2>-fin}
 & \frac{d\,\left<|\widetilde{\pi}(x)|^2\right>}{dx}= -\frac{1}{L_b}\Big[\left<|\widetilde{\pi}(x)|^2\right>+\left<|\widetilde{\gamma}(x)|^2\right>\Big]\ ,\\
\label{<|G|^2>-fin}
 & \frac{d\,\left<|\widetilde{\gamma}(x)|^2\right>}{dx}= -\frac{1}{L_b}\Big[\left<|\widetilde{\pi}(x)|^2\right>+\left<|\widetilde{\gamma}(x)|^2\right>\Big]\ ,\\
\label{<PG*>-fin}
 & \frac{d}{dx}\big<\widetilde{\pi}(x)\widetilde{\gamma}^*(x)\big>=
 -\frac{1}{L_b}\big<\widetilde{\pi}(x)\widetilde{\gamma}^*(x)\big>\ ,\\
\label{<GP*>-fin}
 & \frac{d}{dx}\big<\widetilde{\gamma}(x)\widetilde{\pi}^*(x)\big>=
 -\frac{1}{L_b}\big<\widetilde{\gamma}(x)\widetilde{\pi}^*(x)\big>\ .
\end{align}
\end{subequations}
}
Equations \eqref{<PG*>-fin} and \eqref{<GP*>-fin} with BC \eqref{BC_pi(+)gamma(+)} allow the trivial solutions only,
\begin{equation}\label{<PG*>,<GP*>=0}
  \big<\widetilde{\pi}(x)\widetilde{\gamma}^*(x)\big>= \big<\widetilde{\gamma}(x)\widetilde{\pi}^*(x)\big>\equiv 0\ .
\end{equation}
The first pair of the above equations, namely, \eqref{<|P|^2>-fin} and \eqref{<|G|^2>-fin}, is reduced to the form
\begin{subequations}\label{<|P|^2>pm<|G|^2>}
\begin{align}
\label{<|P|^2>+<|G|^2>}
 & \frac{d}{dx}\Big[\left<|\widetilde{\pi}(x)|^2\right>+\left<|\widetilde{\gamma}(x)|^2\right>\Big]=
\notag\\
 &\phantom{\frac{d}{dx}\left<|\widetilde{\pi}(x)|^2\right>}
 =-\frac{2}{L_b}\Big[\left<|\widetilde{\pi}(x)|^2\right>+\left<|\widetilde{\gamma}(x)|^2\right>\Big]\ ,\\
\label{<|P|^2>-<|G|^2>}
 & \frac{d}{dx}\Big[\left<|\widetilde{\pi}(x)|^2\right>-\left<|\widetilde{\gamma}(x)|^2\right>\Big]= 0\ ,
\end{align}
\end{subequations}
which result in solutions
\begin{subequations}\label{<|P|^2>,<|G|^2>}
\begin{align}
\label{<|P|^2>}
 & \left<|\widetilde{\pi}(x)|^2\right>=
  \frac{|t_+|^2}{2}\left\{\exp\left[\frac{1}{L_b}(L-2x)\right]+1\right\}\ ,\\[6pt]
\label{<|G|^2>}
 & \left<|\widetilde{\gamma}(x)|^2\right>=
  \frac{|t_+|^2}{2}\left\{\exp\left[\frac{1}{L_b}(L-2x)\right]-1\right\}\ .
\end{align}
\end{subequations}
After substituting \eqref{<PG*>,<GP*>=0} and \eqref{<|P|^2>,<|G|^2>} into \eqref{|h(r)|^2-gen(2)}, the expression for the average intensity of the radiation field takes the form
{
\begin{widetext}
\begin{subequations}\label{|h(r)|^2}
\begin{equation}\label{|h(r)|^2-gen(3)}
  \Av{|h(\mathbf{r})|^2} \approx \frac{|t_+|^2}{8}\frac{\sigma^4}{(1+\sigma^2)^2}\left|\overline{Z}_s^{(0)}\right|^2
  \iint\limits_{-\infty}^{\quad\infty}\frac{dqdq'}{(2\pi)^2} \frac{\Psi(q,q',x,z)} {Z_{e}(q)Z_{e} ^*(q')}
  \int_{\mathbb{L}}d\xi\,\Phi(\xi|q)\mathrm{e}^{-i(q-q')\xi}\ ,
\end{equation}
where, to shorten the subsequent formulas, we have introduced the notation
\begin{equation}\label{ToShorten}
  \Phi(\xi|q)=\big[\widetilde{W}^2\big](q-k_{\text{spp}})
  \left[\exp\bigg(\frac{1}{L_b}(L-2\xi)\bigg)+1\right]+
  \big[\widetilde{W}^2\big](q+k_{\text{spp}})\left[\exp\bigg(\frac{1}{L_b}(L-2\xi)\bigg)-1\right]\ .
\end{equation}
\end{subequations}
In polar coordinates the average intensity of the leaking waves at distance $R$ from the origin (placed at the center of segment $\mathbb{L}$) in the direction specified by polar angle $\varphi$ (see Fig.~\ref{fig3}) is given by
\begin{equation}\label{|h(r)|^2-pol(1)}
  \Av{|h(\mathbf{r})|^2}\approx \frac{|t_+|^2}{8}\frac{\sigma^4}{(1+\sigma^2)^2}\left|\overline{Z}_s^{(0)}\right|^2 \int_{\mathbb{L}}d\xi \iint\limits_{-\infty}^{\quad\infty}\frac{dqdq'}{(2\pi)^2}
  \frac{\Psi(q,q',R\cos\varphi,R\sin\varphi)} {Z_{e}(q)Z_{e} ^*(q')}\Phi(\xi|q)\mathrm{e}^{-i(q-q')\xi}\ .
\end{equation}
\begin{figure}[h!!!]
  \scalebox{.6}[.6]{\includegraphics{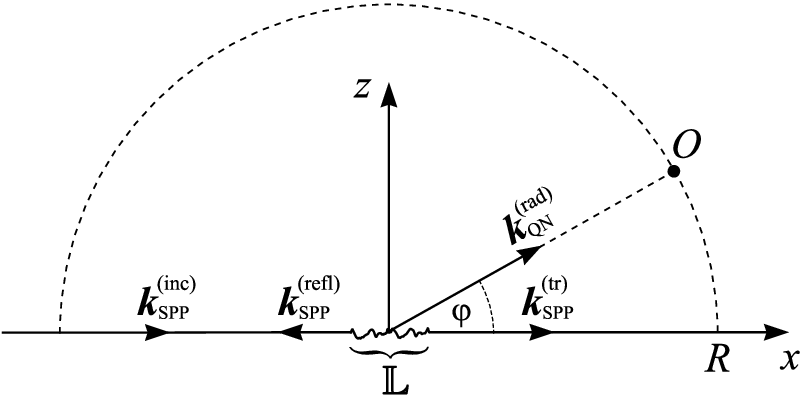}}
  \caption{Schematic view of the scattering of an SPP incident from the left onto rough segment $\mathbb{L}$ of the boundary. Here, $\mathbf{k}_{QN}^{(rad)}$ is the wave vector of the harmonic of the emitted field.}
  \label{fig3}
\end{figure}
For large $R$ the radiation field source, which is the rough segment of the boundary, can be approximately considered as a point-like one. The integrals over $q$ and $q'$ in~\eqref{|h(r)|^2-pol(1)} in this case can be calculated approximately using the stationary phase method. The stationarity points $q_{st}$ of the phase in the exponentials of \eqref{|h(r)|^2-pol(1)} are derived from equality
\begin{align}\label{St_phase-eq}
  R\cos\varphi-\xi=\frac{q_{st} R\sin\varphi}{\sqrt{k^2-q_{st}^2}}\ ,
\end{align}
and can be considered as such provided $k^2-q_{st}^2>0$. Otherwise, these points become the \textit{saddle} points, which are located in the plane of complex variable $q$. From \eqref{St_phase-eq} we get
\begin{align}\label{St_phase-point}
  q_{st}=k\left(\cos\varphi-\frac{\xi}{R}\sin^2\varphi\right) \approx k\cos\varphi\ .
\end{align}
The characteristic ``width'' of the exponential in the vicinity of this point is estimated as
\begin{align}\label{Exp_width}
  (\Delta q)_{ex}\sim \sqrt{\frac{k}{R}}\sin\varphi\ .
\end{align}
If, at sufficiently large $R$, this width is small compared to the width of the Fourier transform of the correlation function $\widetilde{W}$ (which is $\sim 1/r_c$), wherefrom the inequality follows
\begin{align}\label{D_exp<<D_W}
  kr_c\frac{r_c}{R}\sin^2\varphi \ll 1
\end{align}
then
\begin{subequations}\label{|h(r)|^2-Q}
\begin{align}\label{|h(r)|^2-pol(2)}
  \Av{|h(\mathbf{r})|^2}
  \approx & \frac{|t_+|^2}{8}\frac{\sigma^4}{(1+\sigma^2)^2}\left|\overline{Z}_s^{(0)}\right|^2 \frac{kL}{2\pi R} \frac{\sin^2\varphi}{\big|\overline{Z}_s^{(0)}+\sin\varphi\big|^2}
  \bm{Q}(\varphi|L)\ .
\end{align}
Here, for brevity purposes, the notation is used
\begin{align}\label{|h(r)|^2-notat}
  \bm{Q}(\varphi|L)= \big[\widetilde{W}^2\big](k\cos\varphi-k_{\text{spp}})
  \left[\frac{L_b}{2L}\left(\mathrm{e}^{2L/L_b}-1\right)+1\right] + \big[\widetilde{W}^2\big](k\cos\varphi+k_{\text{spp}})
  \left[\frac{L_b}{2L}\left(\mathrm{e}^{2L/L_b}-1\right)-1\right]\ .
\end{align}
\end{subequations}
\end{widetext}
Formula \eqref{|h(r)|^2-pol(2)} allows to track thoroughly the angle distribution of energy radiation of the incident SPP into the upper, dielectric half-space. The dimensionless intensity of the leaking field at distance $R\gg L$ from the rough segment is shown in Fig.~\ref{fig4-new}. It is interesting to notice the pronounced anisotropy of the radiation pattern, which may be effectively tuned by adjusting the inhomogeneity length parameters $r_c$ and $L$, thus enabling the analysis of the role of these parameters in practical measurements.
\begin{figure}[h!]
  \scalebox{.8}[.8]{\includegraphics{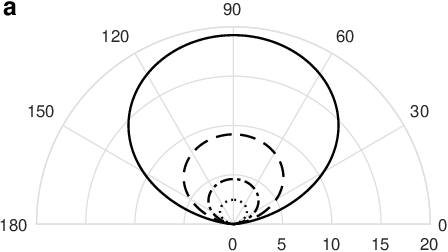}}\qquad\qquad
  \scalebox{.8}[.8]{\includegraphics{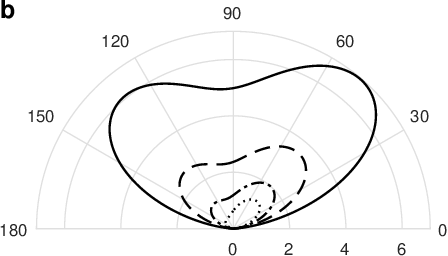}} \\[10pt]
  \scalebox{.8}[.8]{\includegraphics{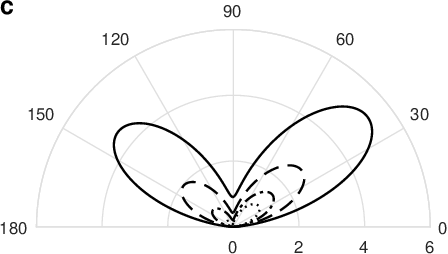}}\qquad\qquad
  \scalebox{.8}[.8]{\includegraphics{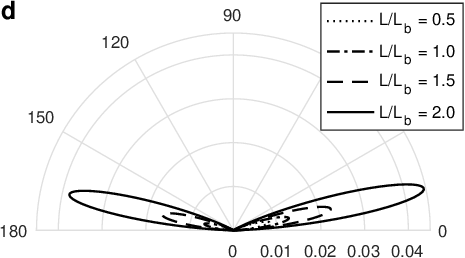}}
  \caption{The scattering patterns of the SPP for different values of parameter $kr_c$: \textbf{\textsf{a}} --- $kr_c=0.1$; \textbf{\textsf{b}} --- $kr_c=1$; \textbf{\textsf{c}} --- $kr_c=1.5$; \textbf{\textsf{d}} --- $kr_c=10$.
  \label{fig4-new}}
\end{figure}

Also, it can be seen from \eqref{|h(r)|^2-Q} that the intensity of the leaking field depends crucially on ratio $L/L_b$. This ratio is characteristic of 1D localization problems where just the length $L_b$ is usually associated with localization of Anderson nature \cite{LifGredPast88}. So, its appearance herein suggests a kind of Anderson localization of the SPP during its propagation across \textit{one-dimensionally} rough surface segment.

The physical quantity \eqref{|h(r)|^2-pol(2)} may seem to be unrelated to strictly 1D wave transport. However, if we return to expression \eqref{|h(r)|^2-gen} for the average intensity, where now the substitution is approved
\begin{equation} \label{K(q,q')}
  \mathcal{R}(q,q') =
  \iint\limits_{-\infty}^{\quad\infty}\frac{dq_1dq_2}{(2\pi)^2}
  \bm{\mathcal{K}}_1(q,q'|q_1,q_2)\bm{\mathcal{K}}_2(q_1,q_2),
\end{equation}
we notice that this formula can well be interpreted as the main term of a series of transitions from the particular purely univariate building block to another one via CP states hidden in the operator $\hat{\mathcal{L}}$ matrix elements. The complete series can be obtained by expanding inverse operators $\big(\openone+\hat{\mathcal{L}}\big)^{-1}$ in \eqref{<R(q)R(q')>} in powers of operator~$\hat{\mathcal{L}}$.

By building blocks we mean the correlators formed by averaging the pair of trial functions $\widetilde{\mathcal{B}}(q_{1,2})$ which are the  univariate objects by definition. Just after the calculation of such ``two-particle'' correlators in coordinate representation there appears the average length~$L_b$, at which one-dimensional eigen-states in a random medium are known to localize. With formula~\eqref{|h(r)|^2-gen}, we actually follow the radiation of not a \textit{free} SPP moving over the surface along the $x$-axis, but rather the radiation of some 1D waves (same SPPs) occupying the \textit{localized states} which are formed dynamically due to the multiple back-scattering of plain 1D harmonics.

The ability of such waves to propagate along the surface with a disordered segment depends, firstly, on the ratio between its intrinsic localization length $L_b$ and the size of the obstacle to be overcome, in our case the rough boundary segment, and secondly, on the efficiency of local transformation of scattered wave from surface to bulk nature. As a measure of transformation efficiency serves the norm of operator $\hat{\mathcal{L}}$ in \eqref{<R(q)R(q')>}, which controls the energy distribution between purely surface and purely bulk states in the one act of such a combined scattering. To determine the net effect of this scattering, either single or multiple one, it is necessary to calculate the scattering coefficients $r_-$ and $t_+$ introduced in trial-field formulas~\eqref{H_0^lrc}.
\vspace{\baselineskip}
%
\section{The search for the scattering coefficients}
\label{Scatt_Coeffs}
%
To determine the reflection and transmission coefficients introduced phenomenologically in Eqs.~\eqref{H_0^<} and \eqref{H_0^>} we will obtain two equations from the conservation law for total energy flux and the boundary conditions at the endpoints of disordered segment~$\mathbb{L}$.
%
\subsection{The flux conservation}
\label{Flow_Cons}
%
In our problem, there can be distinguished four channels for electromagnetic energy flux, which include 1) the incident and reflected SPP waves, see~\eqref{H_0^<}, 2)~the transmitted SPP \eqref{H_0^>}, 3)~the~radiation channel (leaking field $h(\mathbf{r})$), and 4)~the dissipative channel (dissipation directly in the metal). In the impedance model of the interface, the last channel is already taken into account by the real part of the surface impedance ($Z'_s>0$), which sets the SPP dissipative decay length \eqref{L(SPP)_dis}, so we will not dwell on it later on.

For the first and the second channel (we will specify them as SPP channels), the total energy flux at the ``inlet'' and ``outlet'' points of the rough segment can be calculated by integrating over $0<z<\infty$ the energy density $\mathrm{w}_{\text{spp}}=\frac{1}{16\pi }\left(|\mathbf{H}_{\text{spp}}|^2+ |\mathbf{E}_{\text{spp}}|^2\right)$ multiplied by the SPP phase velocity,
\begin{equation}\label{PhaseVel}
  |\mathrm{v}_{\text{spp}}|=c\bigg/\sqrt{1+\Big(\Im\overline{Z}_s^{(0)}\Big)^2}\ ,
\end{equation}
taken with the corresponding sign. In the radiation channel, the phase velocity modulus does not depend on the direction of radiation and equals the speed of light in the uniform upper half-space. Since magnetic field $\mathbf{H}_{\text{spp}}$ has only one nonzero component, ${H_y\equiv H(\mathbf{r})}$, the nonzero electric field components $E_x$ and $ E_z$ are
\begin{equation}\label{ExEz->Hy}
\begin{aligned}
 & E_x=-(i/k)\partial H/\partial z\ ,\\[3pt]
 & E_z=(i/k)\partial H/\partial x\ .
\end{aligned}
\end{equation}
Taking account of  substitutions \eqref{H_0^lrc}, the energy densities of electromagnetic field at the input and the output of interval $\mathbb{L}$ are given by
\begin{widetext}
\begin{subequations}\label{LeftRight_EnDens}
\begin{align}
\label{Left_EnDens}
 & \mathrm{w}_{\text{spp}}^{(\pm)}(-L/2,z)=\frac{1}{16\pi}
\begin{pmatrix}
1\\[3pt]|r_-|^2
\end{pmatrix}
  \left[\Big(1+\big|\overline{Z}_s^{(0)}\big|^2\Big)+
  \left|\sqrt{1-\Big(\overline{Z}_s^{(0)}\Big)^2}\right|^2\right]\exp\Big(2kz\Im\overline{Z}_s^{(0)}\Big) \ , \\
\label{Right_EnDens}
 & \mathrm{w}_{\text{spp}}(L/2,z)=\frac{|t_+|^2}{16\pi}
 \left[\Big(1+\big|\overline{Z}_s^{(0)}\big|^2\Big)+
  \left|\sqrt{1-\Big(\overline{Z}_s^{(0)}\Big)^2}\right|^2\right]\exp\Big(2kz\Im\overline{Z}_s^{(0)}\Big) \ .
\end{align}
\end{subequations}
\end{widetext}
Superscripts $(\pm)$ in \eqref{Left_EnDens} correspond to the SPP propagating in positive and negative directions of the $x$-axis, respectively. Integrating both densities \eqref{LeftRight_EnDens} over $z$ coordinate and multiplying them by velocity \eqref{PhaseVel} taken with corresponding sign, we obtain the expressions for surface wave energy fluxes at both ends of the rough segment, specifically,
\begin{subequations}\label{LrftRight_flows}
 {\allowdisplaybreaks
\begin{align}
\label{Left(+)flow}
 & J^{(+)}(-L/2)=
  \frac{c}{32\pi k\big|\Im\overline{Z}_s^{(0)}\big|\sqrt{1+\Big(\Im\overline{Z}_s^{(0)}\Big)^2}}\times
\notag\\
 & \phantom{J^{(+)}(-L/2)=}
  \times\left[\Big(1+\big|\overline{Z}_s^{(0)}\big|^2\Big)+
  \Big|1-\Big(\overline{Z}_s^{(0)}\Big)^2\Big| \right]\ ,
\\
\label{Left(-)flow}
 & J^{(-)}(-L/2)= -|r_-|^2J^{(+)}(-L/2)\ ,\\
\label{Right(+)flow}
 & J^{(+)}(L/2)= |t_+|^2J^{(+)}(-L/2)\ .
\end{align}
 }
\end{subequations}
If there were no radiative and dissipative scattering channels, then we would obtain the equality well-known in one-dimensional scattering problems, namely,
\begin{align}\label{1D_flow_conserv}
  |r_-|^2+|t_+|^2=1\ ,
\end{align}
which expresses the 1D flux conservation law. However, in the problem considered here, in order to draw up the correct balance of energy flows, it is necessary to account for leaky wave flux radiated into the free (upper) half-space and the flux dissipated within the metal.

Now we will calculate the EM flux radiated outside the metal using time-averaged Poynting vector of the field propagating normally to the semicircle shown in Fig.~\ref{fig3} (actually, the semicylinder of unit length). It is natural to use here a~cylindrical coordinate system with unit vectors $\bm{e}_r$, $\bm{e}_{\phi}$ and $\bm{e}_y$ connected to unit vectors $\bm{e}_x $, $\bm{e}_y$, $\bm{e}_z$ of the original Cartesian system by equalities
\begin{equation}\label{Unit_vect_cyl-Dec}
\begin{aligned}
\begin{cases}
 & \bm{e}_r=\bm{e}_x\cos\varphi+\bm{e}_z\sin\varphi\ ,\\
 & \bm{e}_{\varphi}= -\bm{e}_x\sin\varphi+\bm{e}_z\cos\varphi\ ,\\
 & \bm{e}_y=\bm{e}_y\ .
\end{cases} \\ \
\end{aligned}
\end{equation}
The radial component of the time-averaged (overline symbol) Poynting vector, which determines the density of the energy flux through the cylinder side surface, is equal to
\begin{align}\label{Sr->Ez_Ex}
  \overline{{S}}_r = \frac{c}{8\pi} \Re \big(E_\varphi H_y ^* \big)\ .
\end{align}
Substituting here for $E_\varphi = -E_z\cos\varphi+E_x\sin\varphi$ the electric field components from Eqs.~\eqref{ExEz->Hy}, we obtain
\begin{align}\label{Sr->H_H*}
  \overline{{S}}_r = \frac{c}{8\pi k} \Im
  \left(\frac{\partial H}{\partial x}\cos\varphi+\frac{\partial H}{\partial z}\sin\varphi\right)H^*\ ,
\end{align}
and the expression for the energy flow density averaged over realizations of the rough surface profile
\begin{align}\label{Sr=Real}
  \left<\overline{{S}}_r(\mathbf{r})\right>=\frac{c}{8\pi k}
  \Im\left<\left[\frac{\partial h(\mathbf{r})}{\partial x}\cos\varphi+
  \frac{\partial h(\mathbf{r})}{\partial z}\sin\varphi\right]h^*(\mathbf{r})\right>\ .
\end{align}
The total average flux through the side surface of the half-cylinder is obtained by substituting into field \eqref{Full_solution_h(r)} the values $x=R\cos\varphi,\ z=R\sin\varphi$ and integrating the resulting expression over interval $0<\varphi<\pi$.

Let us denote the flows defined by the first and the second terms in \eqref{Sr=Real} by symbols $\Av{J^{(rad)}_{1}}$ and $\Av{J^{(rad)}_ {2}}$, respectively. The explicit form of the first of these fluxes~is

\begin{widetext}
\begin{equation}\label{Rad_term-1}
  \left<J^{(rad)}_{1}\right> = \frac{cR}{8k}\Re\int\limits_0^{\pi}d\varphi\cos\varphi
  \iint\limits_{-\infty}^{\quad\infty}\frac{qdqdq'}{(2\pi)^2}
  \Psi(q,q',R\cos\varphi,R\sin\varphi) \mathcal{R}(q,q')\ .
\end{equation}
\end{widetext}
The double integral over $q$ and $q'$ here coincides up to the integrand factor $q$ with the analogous integral in~\eqref{|h(r)|^2-gen}. Therefore, it can be calculated in the same way (by means of the stationary phase method) that was used to obtain \eqref{|h(r)|^2-pol(2)}. The difference between expression \eqref{Rad_term-1} and the corresponding part of \eqref{|h(r)|^2-pol(2)} is only in the appearance of additional factors $cR/8k$ and $q_0=k\cos\varphi $, as well as in additional integration over angle $\varphi$. Taking this into account we get
\begin{align}\label{Rad_term-1_fin}
 & \Av{J^{(rad)}_{1}} = \frac{|t_+|^2}{8}\frac{\sigma^4}{(1+\sigma^2)^2}\left|\overline{Z}_s^{(0)}\right|^2 
\notag\displaybreak[0]\\
 & \phantom{\Av{J^{(rad)}_{1}} =}
 \times\frac{ckL}{32\pi} \int\limits_0^{\pi}d\varphi\frac{\sin^2\varphi\cos^2\varphi}{\Big|\overline{Z}_s^{(0)}+
  \sin\varphi\Big|^2}\, \bm{Q}(\varphi|L)\ .
\end{align}

The expression for the second term in radiative flux~\eqref{Sr=Real}, denoted by $\Av{J^{(rad)}_{2}}$, differs from \eqref{Rad_term-1} in that factor $q $ in it must be replaced with $\sqrt{k^2-q^2}$, and factor $\cos^2\varphi$ in the integral over $\varphi$ be substituted for $\sin^2\varphi$. At the stationary phase point~\eqref{St_phase-point} the equality $\sqrt{k^2-q^2}\approx k\sin\varphi$ is fulfilled, which gives rise to the resulting formula
\begin{align}\label{Rad_term-2_fin}
 \Av{J^{(rad)}_{2}} =&  \frac{|t_+|^2}{8}\frac{\sigma^4}{(1+\sigma^2)^2}\left|\overline{Z}_s^{(0)}\right|^2  \times
\notag\displaybreak[0]\\
 & \times\frac{ckL}{32\pi}
  \int\limits_0^{\pi}d\varphi\frac{\sin^4\varphi}{\Big|\overline{Z}_s^{(0)}+\sin\varphi\Big|^2} \, \bm{Q}(\varphi|L)\ .
\end{align}
The flow balance is now written as
\begin{align}\label{Flow_balance}
  J^{(+)}(-L/2)+J^{(-)}(-L/2)=J^{(+)}(L/2)+\Av{J^{(rad)}}\ ,
\end{align}
which, obviously, does not reduce to equality \eqref{1D_flow_conserv} typical for one-dimensional non-dissipative systems. Dividing \eqref{Flow_balance} by $J^{(+)}(-L/2)$, we get the flow balance equation in the following form,
\begin{align}\label{Flow_balance-fin}
  1-|r_-|^2=|t_+|^2\left(1+\Av{I^{(rad)}}\right)\ .\\[-.7\baselineskip] \notag
\end{align}
Here $\Av{I^{(rad)}}$ is the normalized radiative flux, which in the limiting case $\|\hat{\mathcal{L}}\|\ll 1$ is represented by expression as follows,
\begin{widetext}
\begin{align}\label{Norm_rad_flow}
  \Av{I^{(rad)}} & =  \frac{\sigma^4}{8(1+\sigma^2)^2}k^2L
  \frac{\left|\overline{Z}_s^{(0)}\right|^2\left|\Im\overline{Z}_s^{(0)}\right|\sqrt{1+\Big(\Im\overline{Z}_s^{(0)}\Big)^2}}
  {\left(1+\Big|\overline{Z}_s^{(0)}\Big|^2\right)+\left|1-\Big(\overline{Z}_s^{(0)}\Big)^2\right|}
  \int\limits_0^{\pi}d\varphi\frac{\sin^2\varphi}{\big|\zeta_0+\sin\varphi\big|^2} \, \bm{Q}(\varphi|L)\ .
\end{align}
\end{widetext}
%
\subsubsection{On the possibility of tuning the radiative flux over the polar angle}
\label{I(rad)-anisotropy}
%
Correlation functions $\widetilde{W}(k_{\text{spp}}\pm k\cos\varphi)$ entering into \eqref{|h(r)|^2-pol(2)} and \eqref{Norm_rad_flow} trough function $\bm{Q}(\varphi|L)$ are either smooth or sharp functions of angle $\varphi$ over its variation interval, depending on the relationship between correlation radius $r_c$ and SPP wavelength ${\lambda_{\text{spp}}\sim k^{-1}_{\text{spp}}\sim k^{-1}}$. If $kr_c\ll 1$, both of these functions differ slightly from $\widetilde{W}(0)$, and the dependence of the radiated energy density on angle $\varphi$ is determined by factor
\begin{align}\label{Phi_distrib-smooth}
  \mathcal{E}(\varphi)=\frac{\sin^2\varphi}{\big|\zeta_0+\sin\varphi\big|^2}
\end{align}
under the integral in \eqref{Norm_rad_flow}.

If $kr_c\gg 1$, then both of Fourier transforms of the correlation function in \eqref{|h(r)|^2-pol(2)} are small compared to the value of $\widetilde{ W}(0)\sim r_c$, since for almost any angle $\varphi$ the estimate holds true $|k_{\text{spp}}\pm k\cos\varphi|\sim k$. This implies that it is practically impossible to affect significantly the \textit{value} of the energy density of the radiation field as well as its flux. Only their \textit{angular structure} can be influenced noticeably, and the main tuning parameter here is the dimensionless quantity $kr_c$. Figure~\ref{fig4-new} clearly demonstrates the aforementioned transition from quasi-isotropic to sharply anisotropic angular distribution of radiation with increasing the roughness correlation length~$r_c$.
%
\subsubsection{The effect of boundary conditions at the rough segment endpoints}
\label{End_BCs}
%
The flow balance \eqref{Flow_balance-fin} establishes a relationship between scattering coefficients $r_-$ and $t_+$, but it is not sufficient to determine their specific values. To obtain one more equation connecting $r_-$ and $t_+$ we turn to the BCs at the end points of disordered segment $\mathbb{L}$.

According to our model, the magnetic field of the wave is represented as a sum of trial field $ H_0(\mathbf{r}) $ and radiation field $h(\mathbf{r})$. At the endpoints of $\mathbb{L}$ the latter field tends to zero (see paragraph following Eq.~\eqref{Helmholtz_on_all_x}), and, therefore, at these points it is necessary to match the ``external'' (with respect to $\mathbb{L}$) SPP fields and the ``internal'' trial field $H_0^{(in)}(\mathbf{r})$, which we model by expression~\eqref{in_trial field}. The matching should be performed for $z=0$ only, since for $z>0$ on vertical axes $x=\pm L/2$ radiation field $h(\mathbf{r})$ is not equal exactly to zero. In other words, at the endpoints of $\mathbb{L}$, directly at $z=0$, fields \eqref{H_0^<} and \eqref{H_0^>} from the outer sides of the rough segment and the ``internal'' SPP field \eqref{in_trial field} must be precisely matched.

The latter field we represent in the form \eqref{B(x)_WS}. For functions $\pi(x)$ and $\gamma(x)$, the BCs \eqref{BC_pi(+)gamma(+)} and~\eqref{BC_pi(-)gamma(-)} were obtained, from which until now the conditions \eqref{BC_pi(+)gamma(+)} only were utilized. Note that BC sets \eqref{BC_pi(+)gamma(+)} and \eqref{BC_pi(-)gamma(-)} were obtained just by means of matching the ``inner'' trial field \eqref{H_0(in)} with complementary ``outer'' SPP fields \eqref{H_0^lrc} at~${z=0}$.
Formally, one could relate both of the above BCs, \eqref{BC_pi(+)gamma(+)} and~\eqref{BC_pi(-)gamma(-)}, by directly solving equations~\eqref{pi_gamma-eqs}, with solely one of them taken as a set of ``initial'' values. Yet, this is hard to implement in practice due to the random nature of functions $\eta(x)$ and $\zeta_{\pm}(x)$. So, we will obtain the connection between two-side BCs on the interval~$\mathbb{L}$ without solving directly the dynamic equations on it.

This, however, can be done not fully exactly but with statistical accuracy only, in the so-called ``correlation approximation'', when instead of the exact random function its statistically averaged version is utilized. Applying the averaging procedure outlined in section \ref{ScattDiagr} to equations \eqref{pi_gamma-eqs} we get
\begin{subequations}\label{<pi-><gamma->-fin}
\begin{align}
\label{<pi->-fin}
 & \Av{\pi(x)}= t_+\mathrm{e}^{-ik_{\text{spp}}L/2}
\notag\\
 & \phantom{\Av{\pi(x)}}
 = \exp\Bigg[-\frac{1}{2}\left(\frac{1}{L_f} -\frac{1}{L_b}\right) \big(L/2-x\big)\Bigg]\ ,\\
\label{<gamma->-fin}
 & \Av{\gamma(x)}\equiv 0\ .
\end{align}
\end{subequations}
Equalities \eqref{<pi-><gamma->-fin} allow to connect the values of mean field $\mathcal{B}(x)$ on the inner sides of both ends of the rough segment with regular ``outer'' fields \eqref{H_0^<} and \eqref{H_0^>}, thus obtaining the missing relationship between coefficients $r_-$ and~$t_+$.

For $x=L/2$, the BC at the right end of $\mathbb{L}$ has already been used when performing statistical averaging. Now, with the same functions \eqref{<pi-><gamma->-fin} we must also satisfy the matching condition at point ${x=-L/2}$. This results in relationship
\begin{align}\label{Left_BC}
  1+r_-= t_+ \exp\left[-\left(\frac{1}{L_f}-\frac{1}{L_b}\right)\frac{L}{2}\right]\mathrm{e}^{-ik_{\text{spp}}L}\ ,
\end{align}
which can be considered asymptotically valid in various limiting cases with respect to the length of the rough segment. If ${L\ll L_{f,b}}$ (so called ``quasi-ballistic'' regime of the wave transport), from \eqref{Left_BC} we get the approximate equality
\begin{equation}\label{Ball_limit}
  1+r_-\approx t_+ \mathrm{e}^{-ik_{\text{spp}}L}\ .
\end{equation}
In the opposite limiting case $L\gg L_{f,b}$, if $L_b$ is not equal to $L_f$ exactly, the incident SPP can penetrate into segment $\mathbb{L}$ no more than at a distance $\sim L_b$, which is associated with the length of (one-dimensional) Anderson localization. The right-hand side of \eqref{Left_BC} then becomes negligibly small as compared to unity, and with asymptotic accuracy we arrive at equality
\begin{equation}\label{r=-1}
  r_-\approx -1\ ,
\end{equation}
which implies that the incident SPP is almost completely back-reflected from the rough boundary segment.

Formula \eqref{Left_BC} suggests one more interesting result. The exponent in the square brackets of \eqref{Left_BC} can be made small as compared to unity both in quasi-ballistic limit $L\ll L_{f,b}$ and in localization regime $L_b\ll L$, provided that the difference of reciprocal scattering lengths $L_f$ and $L_b$ is small as compared to the inverse length of the rough segment. Taking account of the exact value of scattering lengths calculated in subsection~\ref{Pots_corrs}, this is possible for sufficiently small correlation length, when parameter $kr_c$ becomes small as compared to unity.

In this particular case, as one of conceivable explanations for the lack of exponential localization one can assume that within the disordered segment the incident SPP does not make a distinction between multiple ``forward'' and ``backward'' scattering. As a~result, its dynamics within the segment becomes effectively isotropic, and instead of occupying the Anderson-localized state pinned to the left, ``input'' edge of $\mathbb{L}$ it fills quasi-freely (diffusionally) the entire segment, thus forming within its framework an effective Fabry-Perot resonator. The resonant states in the resonator may not exactly coincide with the states in ``shore'' areas $|x|> L/2$, although they may to some extent overlap with the latter ones.

Such a discrepancy between the ``outer'' and the ``inner'' (with respect to $\mathbb{L}$) states, which arises due to specific ``index mismatch'' in a finite one-dimensionally disordered open-side system, should significantly affect both scattering coefficients $r_-$ and $t_+$, and the intensity of the radiative component of the scattered field. The detailed analysis of this mechanism of SPP scattering by a randomly rough surface segment is the task for further analysis.
%
\section{Discussion of the results}
\label{Conclusions}
%
To conclude, in this paper we have developed a detailed theory of surface plasmon-polariton scattering by a~finite segment of metal-vacuum interface with randomly distributed on it surface asperities. The geometric non-uniformity of the interface results in the scattering of initial purely surface wave into both surface harmonics of SPP type and a bunch of bulk wave harmonics propagating freely at different angles in the half-space above the metal surface.

The scattered harmonics, both of SPP and of leaky types, are shown to be intimately coupled. For the measure of their interconnection we suggest the appropriate criterion, namely, the value of Hilbert-Schmidt norm of the operator describing the intermediate scattering states, the so called composite plasmons. These intermediate wave excitations represent a~specially composed superposition of purely SPP waves and of a~bundle of bulk harmonics through which the energy is emitted from the metal surface into vacuum or dielectric volume.

The attempt to explain the experimentally observed sharp peaks of the local field in random metal-dielectric composites by Anderson localization of surface plasmons was undertaken in \cite{MARADUDIN1994302}, but later on this interpretation was declined \cite{MaraduFNT2010}. In our paper \cite{TarStadKvit23}, a different mechanism was proposed to explain the emission of SPP waves from the metal boundary into the bulk of a dielectric, which reduces to the excitation of composite plasmons.

The perturbation of wave-guiding surface was considered in \cite{TarStadKvit23} in the form of coordinate-dependent surface impedance. In the present study the impedance is assumed to be locally unchanged, whereas as a variable quantity is chosen the relief of the interface. Such a formulation of the problem is physically justified for uneven surfaces whose local curvature is small in comparison with the inverse skin layer thickness in the metal~\cite{Maradudin1995}.

We have shown that if the conditions for introducing the local boundary impedance are satisfied, the effect of geometrical perturbation of the interface can be described in terms of \textit{effective} (nonlocal and nonhomogeneous, in the general case) surface impedance. The dependence of the effective impedance on the coordinates and the degree of the nonlocality is determined by the value of the asperity sharpness, i.\,e., by the derivative of the interface profile.

If the boundary is non-uniform along one coordinate only and sufficiently extended in the direction of propagation of the incident SPP, the scattering can be considered as having two-channel nature. The first is the SPP channel, in which the scattered waves propagate along one axis only, having all the attributes of the surface polariton. In the second channel, the two-dimensional propagation and radiative emission of the initially surface wave does occur. In the purely one-dimensional channel, the Anderson localization of SPP harmonics, which arises due to the interference during their multiple scattering along the propagation path, is manifested significantly. But since both of the above channels are rigidly coupled, the localization in the one-dimensional surface channel also crucially affects the level of SPP radiation into the free dielectric half-space.

For mathematically rigorous description of the interference-induced localization in the SPP channel, we have developed a~method for effective one-dimensionalization of the initially posed two-dimensional scattering problem. Within this method, the solution of Helmholtz equation is sought in the form of a~sum of the SPP-type trial wave, which is dealt with as a~strictly one-dimensional object, and the additives in the form of two-dimensional leaky waves excited at the rough boundary segment, which are generated by the trial field.

If the 1D trial wave is localized within the rough segment, then for relatively large correlation length of the roughness, $r_c\gg k^{-1}$, the behavior of the incident SPP is practically the same as for any monochromatic wave incident from a~homogeneous medium onto a randomly layered half-space \cite{Papanic94}. Specifically, the incident plasmon-polariton should be almost completely reflected in the backward direction, and only a relatively small (yet not of exponential smallness!) part of its energy is to be sharply anisotropically scattered into the free dielectric half-space. Only an exponentially small (in parameter $\sim\exp(-L/L_{loc})$) part of the incident SPP can pass through the disordered segment of the boundary in the form of a purely surface wave.

In the opposite case $r_c\ll k^{-1}$, the ``forward'' and the ``backward'' scattering lengths of the trial polariton coincide with each other with parametric accuracy, and there is no simple relationship between the reflection and the transmission coefficients in the correlation approximation. We can expect that with a relatively small longitudinal inhomogeneity of the interface, its rough segment may be considered as an open section of a 1D waveguide, dielectric filling of which does not coincide with the fillings of its side ``shores''. In such a piecewise inhomogeneous near-surface waveguide, the Fabry-Perot-type resonances have to occur, due to which the passage of the SPP across the rough interface segment may significantly increase in comparison to the case where the trial wave is completely localized due to its stochastic scattering by the boundary relief. We consider the detailed analysis of this nontrivial issue as an outlook for the forthcoming study.
\section*{Acknowledgement}
Yu.\,T. gratefully acknowledges the support from the National Research Foundation of Ukraine, Project No.~2020.02/0149 ``Quantum phenomena in the interaction of
electromagnetic waves with solid-state nanostructures''.

\begin{widetext}
\appendix
%
\section{Calculation of correlators (\ref{K_1K_2-corrs})}
\label{K_1K_2-calc}
%
Correlators $\bm{\mathcal{K}}_1(q,q'|q_1,q_2)$ and $\bm{\mathcal{K}}_2(q_1,q_2)$ from \eqref{K_1K_2-corrs} can be presented in the form
\begin{subequations}\label{K_1K_2-full}
\begin{align}\label{K_1-full}
  \bm{\mathcal{K}}_1(q,q'|q_1,q_2)=
  \frac{1}{Z_{e}(q)Z_{e} ^*(q')}
  \Big<\Delta\widetilde{\overline{Z}}_s(q-q_1)\Delta\widetilde{\overline{Z}}^*_s(q'-q_2)\Big>
\end{align}
and, correspondingly,
\begin{align}\label{K_2-full}
  \bm{\mathcal{K}}_2(q_1,q_2)=&
  \iint_{\mathbb{L}}dx_1dx_2\mathrm{e}^{-iq_1x_1+iq_2x_2}
  \left[\Av{\pi(x_1)\pi^*(x_2)}\mathrm{e}^{ik_{\text{spp}}(x_1-x_2)}+
  \Av{\gamma(x_1)\gamma^*(x_2)}\mathrm{e}^{-ik_{\text{spp}}(x_1-x_2)} \right. 
 \notag\\
 &\phantom{\iint_{\mathbb{L}}dx_1dx_2\mathrm{e}^{-iq_1x_1+iq_2x_2}\quad}
  + i\Av{\pi(x_1)\gamma^*(x_2)}\mathrm{e}^{ik_{\text{spp}}(x_1+x_2)}-
  \left.i\Av{\gamma(x_1)\pi^*(x_2)}\mathrm{e}^{-ik_{\text{spp}}(x_1+x_2)}\right]\ .
\end{align}
\end{subequations}
When deriving expression \eqref{K_2-full} we have neglected the difference between $k_{\text{spp}}$ and $k^*_{\text{spp}}$, that is, we have considered this wave parameter as real-valued. This is admissible, strictly speaking, if the SPP dissipation length in the metal exceeds significantly the rough segment length, i.\,e., if the inequality holds true $L^{\scriptscriptstyle{(\text{SPP})}}_{dis}\gg L$.

Consider now the correlation function from \eqref{K_1-full}, which may be written as
\begin{align}\label{K_1-sub}
 \Big<\Delta\widetilde{\overline{Z}}_s(\kappa)\Delta\widetilde{\overline{Z}}^*_s(\kappa')\Big> = &
  \iint_{\mathbb{L}}dx_1dx_2 \mathrm{e}^{-i\kappa x_1+i\kappa' x_2}
  \Big<\Delta\overline{Z}_s(x_1)\Delta\overline{Z}^*_s(x_2)\Big> = \left|\overline{Z}_s^{(0)}\right|^2
  \iint_{\mathbb{L}}dx_1dx_2 \mathrm{e}^{-i\kappa x_1+i\kappa'x_2}
 \notag\\
 & \times \Bigg< \left[\frac{1}{\sqrt{1+\Delta\left[\xi'(x_1)\right]^2\big/(1+\sigma^2)}}-1\right]
  \left[\frac{1}{\sqrt{1+\Delta\left[\xi'(x_2)\right]^2\big/(1+\sigma^2)}}-1\right] \Bigg> \ .
\end{align}
The correlator under double integral in \eqref{K_1-sub} can be calculated approximately if we consider the ratio ${\Delta\left[\xi'(x_{1,2})\right]^2\big/(1+\sigma^2 )}$ under the root signs to be small compared to unity and random function $\xi'(x)$ to be Gaussian distributed. After expanding the square roots in \eqref{K_1-sub} to the first terms after unity we have to calculate the auxiliary correlator
\begin{align}\label{<Auxil-1>}
  \mathcal{K}_{aux}(x_1-x_2) = 2\sigma^4\AV{\Delta\left[\xi'(x_{1})\right]^2\Delta\left[\xi'(x_{2})\right]^2}W^2(x_1-x_2) \ .
\end{align}
Correlator \eqref{K_1-sub} is then written as
\begin{align}\label{K_1-sub(2)}
  \Big<\Delta\widetilde{\overline{Z}}_s(\kappa)\Delta\widetilde{\overline{Z}}^*_s(\kappa')\Big>=
  \frac{\left|\overline{Z}_s^{(0)}\right|^2 \sigma^4}{2(1+\sigma^2)^2} \iint_{\mathbb{L}}dx_1dx_2 \mathrm{e}^{-i\kappa x_1+i\kappa'x_2}W^2(x_1-x_2) \ ,
\end{align}
and correlator \eqref{K_1-full}, correspondingly, in the form
\begin{equation}\label{K_1-full(2)}
  \bm{\mathcal{K}}_1(q,q'|q_1,q_2)=
 \frac{\left|\overline{Z}_s^{(0)}\right|^2 \sigma^4}{2(1+\sigma^2)^2Z_{e}(q)Z_{e} ^*(q')}
 \iint_{\mathbb{L}}dx_1dx_2 \mathrm{e}^{-i(q-q_1)x_1+i(q'-q_2)x_2}W^2(x_1-x_2) \ .
\end{equation}
Function $W^2(x_1-x_2)$ can be represented as Fourier integral
\begin{equation}\label{W^2-Fourier(x)}
  W^2(x_1-x_2)= \int\limits_{-\infty}^{\infty}\frac{dQ}{2\pi}\left[\widetilde{W}^2\right](Q)\mathrm{e}^{iQ(x_1-x_2)}\ ,
\end{equation}
where
\begin{align}\label{W^2-Fourier(Q)}
  \left[\widetilde{W}^2\right](Q)=\int\limits_{-\infty}^{\infty}\frac{d\kappa}{2\pi}
  \widetilde{W}(\kappa)\widetilde{W}(Q-\kappa)\ .
\end{align}
%
Substituting \eqref{W^2-Fourier(x)} into \eqref{K_1-full(2)} and calculating the integrals over $x_1$ and $x_2$ we get
\begin{align}\label{K_1-full(3)}
 & \bm{\mathcal{K}}_1(q,q'|q_1,q_2)=
 \Delta_L(q-q_1-q'+q_2)\frac{\left|\overline{Z}_s^{(0)}\right|^2 \sigma^4}{2(1+\sigma^2)^2Z_{e}(q)Z_{e} ^*(q')}  \int\limits_{-\infty}^{\infty}\frac{d\kappa}{2\pi}\widetilde{W}(\kappa)\widetilde{W}(q-q_1-\kappa) \ .
\end{align}

The integrals over $q_1$ and $q_2$ in \eqref{|h(r)|^2-gen} can be calculated using \eqref{K_1-full(3)}, \eqref{K_2-full} and the following property of $\Delta_L$-function,
\begin{equation}\label{Delta_L-prop_1}
  \int\limits_{-\infty}^{\infty}\frac{dq}{2\pi}\mathrm{e}^{iqx}\Delta_L(q-Q)=
  \mathrm{e}^{iQx}\theta(L/2-|x|)\ .
\end{equation}
The result of such an integration in formula \eqref{K(q,q')} is
%
\begin{align}\label{Int_q1q2_K1K2-1}
 & \iint\limits_{-\infty}^{\quad\infty} \frac{dq_1dq_2}{(2\pi)^2}\bm{\mathcal{K}}_1(q,q'|q_1,q_2) \bm{\mathcal{K}}_2(q_1,q_2)=
\frac{\left|\overline{Z}_s^{(0)}\right|^2 \sigma^4}{2(1+\sigma^2)^2Z_{e}(q)Z_{e} ^*(q')}
\notag\\
 & \qquad \times \iint_{\mathbb{L}}dx_1dx_2 W^2(x_1-x_2)\mathrm{e}^{-iqx_1+iq'x_2}
  \Big[\Av{\pi(x_1)\pi^*(x_2)}\mathrm{e}^{ik_{\text{spp}}(x_1-x_2)}+
  \Av{\gamma(x_1)\gamma^*(x_2)}\mathrm{e}^{-ik_{\text{spp}}(x_1-x_2)}
\notag\\
 & \phantom{\iint_{\mathbb{L}}dx_1dx_2 W^2(x_1-x_2)\mathrm{e}^{-iqx_1+iq'x_2}+\pi(x_1)}
  +i\Av{\pi(x_1)\gamma^*(x_2)}\mathrm{e}^{ik_{\text{spp}}(x_1+x_2)}-
   i\Av{\gamma(x_1)\pi^*(x_2)}\mathrm{e}^{-ik_{\text{spp}}(x_1+x_2)}\Big]\ .
\end{align}
Changing the integration variables $x_1$ and $x_2$ to $x_1-x_2$ and $(x_1+x_2)/2$ and taking into account that functions $\pi(x)$ and $\gamma(x)$ are smooth compared to correlation factor $W^2(x)$ we get
\begin{align}\label{Int_q1q2_K1K2-3}
  \iint\limits_{-\infty}^{\quad\infty} \frac{dq_1dq_2}{(2\pi)^2} \bm{\mathcal{K}}_1(q,q'|q_1,q_2)\bm{\mathcal{K}}_2(q_1,q_2) = &
 \frac{\left|\overline{Z}_s^{(0)}\right|^2 \sigma^4}{2(1+\sigma^2)^2Z_{e}(q)Z_{e} ^*(q')}
\notag\\
  & \times \int_{\mathbb{L}}d\xi
  \left\{\Av{|\pi(\xi)|^2}\big[\widetilde{W}^2\big](q-k_{\text{spp}})+
  \Av{|\gamma(\xi)|^2}\big[\widetilde{W}^2\big](q+k_{\text{spp}})\right.
\notag\\
 & \qquad +i\Av{\pi(\xi)\gamma^*(\xi)}\big[\widetilde{W}^2\big](q-k_{\text{spp}})-
  \left.i\Av{\gamma(\xi)\pi^*(\xi)}\big[\widetilde{W}^2\big](q+k_{\text{spp}})\right\}
  \mathrm{e}^{-i(q-q')\xi}\ .
\end{align}
\end{widetext}
%
%
%

\end{document}